\documentclass[12pt]{article}

\usepackage[utf8]{inputenc}
\usepackage[T1]{fontenc}  

\usepackage[a4paper,top=1.6in,bottom=2in,left=1.25in,right=1.25in]{geometry}
\usepackage{setspace}
\usepackage{amsmath, amsthm, amssymb, mathtools}
\usepackage{bm}           
\usepackage{bbm}          
\usepackage{mathrsfs}     
\usepackage{bigints}      
\usepackage{commath}      
\usepackage{accents}      
\usepackage{arydshln}     

\usepackage{tabularx}
\usepackage{booktabs}
\usepackage{array}

\newlength{\dhatheight}

\usepackage{titlesec}
\titleformat*{\section}{\Large\bfseries}
\titleformat*{\subsection}{\large\bfseries}
\titleformat*{\subsubsection}{\large\bfseries}
\usepackage{enumerate}
\usepackage{verbatim}
\usepackage{listings}
\usepackage{xcolor}

\usepackage{float}
\usepackage{booktabs}
\usepackage{subcaption}
\usepackage{adjustbox}
\usepackage{multirow}
\usepackage{rotating}
\usepackage{colortbl}
\usepackage{lscape}

\usepackage{natbib}
\setcitestyle{authoryear}
\usepackage{hyperref}
\hypersetup{
    colorlinks=true,
    linkcolor=blue,
    citecolor=blue,
    urlcolor=blue,
    hypertexnames=true
}

\newtheorem{assumption}{Assumption}[section]
\newtheorem{theorem}{Theorem}[section]

\numberwithin{equation}{section}

\def\equationautorefname~#1\null{equation~(#1)\null}

\usepackage[title]{appendix}

\usepackage{authblk}

\usepackage{algorithm}
\usepackage{algpseudocode}

\usepackage{tikz}
\usetikzlibrary{decorations.pathmorphing, patterns, shapes, arrows.meta, calc, fit, positioning}
\tikzset{
    -Latex,auto,node distance =1 cm and 1 cm,semithick,
    state/.style ={ellipse, draw, minimum width = 0.7 cm},
    point/.style = {circle, draw, inner sep=0.04cm,fill,node contents={}},
    bidirected/.style={Latex-Latex,dashed},
    el/.style = {inner sep=2pt, align=left, sloped},
    rct/.style = {draw}
}

\makeatletter
\patchcmd{\maketitle}
  {\renewcommand\thefootnote{\@fnsymbol\c@footnote}}
  {\renewcommand\thefootnote{\alph{footnote}}}
  {}{}
\makeatother

\renewcommand{\thefootnote}{\fnsymbol{footnote}}

\begin{document}

\begin{center}
{\LARGE Local projections identify the same policy counterfactuals as empirical and structural models} \\[1.25em]
{\large Endong Wang\footnote[1]{%
This paper is based on my job market paper \emph{``Structural Counterfactual Analysis in Macroeconomics: Theory and Inference''} completed at McGill University. 
I am indebted to my advisors Jean-Marie Dufour, Victoria Zinde-Walsh, and Russell Davidson 
for their guidance and support during my Ph.D.\ studies. 
I also thank Francisco Alvarez-Cuadrado, Santiago Camara, Rui Castro, Christian Conrad, Valentina Corradi, John Galbraith, Silvia Gon\c{c}alves,  Christian Gourieroux, Dimitris Korobilis,  Ricardo P. C. Nunes, Marcelo Rodrigues dos Santos, Shuping Shi, Carsten Trenkler, and Ke-Li Xu 
for helpful comments and suggestions. 
This work was supported by the Fonds de recherche du Qu\'ebec -- Soci\'et\'e et culture (FRQSC)
and the research funds provided by the University of Mannheim.}%
\footnote[2]{Department of Economics, University of Mannheim, L7, 3–5, Room 124, Mannheim, Germany, 68161. 
TEL: +49 621-181-1879; e-mail: \href{mailto:endong.wang@uni-mannheim.de}{endong.wang@uni-mannheim.de}. 
Web: \url{http://www.endongwang.com}.}} \\[0.25em]
{\today}
\end{center}

\renewcommand{\thefootnote}{\arabic{footnote}}
\setcounter{footnote}{0}

\onehalfspacing
\pagenumbering{arabic}
\setcounter{section}{0}
\setcounter{page}{1}

\begin{abstract}
We study policy counterfactuals that impose path restrictions on a policy instrument over a finite window. Under a sequential intervention design, we
define two counterfactual objects, policy-peg impulse responses and policy-path effects, and we provide a novel local projection identification method. Under policy invariance and a linear moving average envelope, the local projection estimands coincide with the counterfactual outcomes implied by empirical vector autoregressions and linearized forward looking structural models, and the counterfactual outcomes are fully characterized by the relevant impulse responses. We also provide local projection identification of both counterfactual objects under an one-shot intervention design. In the empirical applications, we quantify the propagation of an oil-supply news shock under interest-rate pegs and study alternative liftoff paths during the post-pandemic tightening episode.

\end{abstract}


\section{Introduction}

A central task in macroeconomic policy analysis is to quantify how
macroeconomic outcomes would evolve under counterfactual interventions that
constrain the policy instrument over a finite window. Two classes of exercises
are standard. The first imposes a temporary peg of the policy instrument
following a shock of interest and asks how the restriction reshapes
propagation, thereby isolating the role of systematic policy in the
transmission of the shock, e.g., \cite{sims1995does,bernanke1997systematic}. The
second replaces the realized policy path by an alternative path. When the baseline corresponds to a historical episode,
such path replacements support retrospective assessments of policy conduct.\footnote{See, e.g.,
\cite{sims1995does,waggoner1999conditional,leeper2003modest,hamilton2004comment,sims2006does,kilian2011does,laseena2011anticipated,del2013dsge,banbura2015conditional,davis2019monetary,antolin2021structural,christian2024evaluating}.}
We refer to the associated counterfactual objects as policy-peg impulse responses
and policy-path effects.

These policy counterfactuals are typically implemented within a fully specified
structural vector-autoregression (SVAR) or linearized structural model, coupled with a recursion that enforces the restriction through
period-by-period interventions in the policy shock, as in
\cite{sims1995does,sims2006does,laseena2011anticipated,del2013dsge}. Recent work adopts a sufficient-statistics perspective, expressing many policy counterfactual questions in terms of impulse responses rather than a fully specified model, see \cite{chetty2009sufficient,barnichon2023sufficient,beraja2023semistructural,mckay2023can,caravello2024evaluating,barnichon2026policy}. Existing formulations, however, largely emphasize rule counterfactuals, where alternative paths arise endogenously from changes in the systematic policy rule, whereas many applied exercises instead take a policy path restriction as primitive. Moreover, although impulse responses coincide in population across VAR and local projections (LP) under certain conditions, see \cite{dufour1998short,jorda2005estimation,wolf2021same}, it remains unclear whether path-restriction counterfactuals admit an analogous regression estimand, and hence whether the VAR–LP equivalence extends beyond impulse-response objects.

This paper develops a local projection framework for policy counterfactuals. The
framework represents counterfactual responses as regression objects. We focus on
a sequential intervention design in which the restriction is implemented by
period-by-period policy-shock interventions that are not anticipated by private
agents, paralleling the conventional SVAR construction \cite{sims1995does,sims2006does}.
We also discuss LP identification with the one-shot counterfactual design of \cite{mckay2023can}, in which the entire
path restriction is implemented at the event date. Throughout, our emphasis is on identification and on
a regression characterization of macro counterfactuals.

We make three contributions. First, under the sequential design, we define
policy-peg impulse responses and policy-path effects in a generic expectations
representation that does not impose linearity or a specific parametric model.
We then derive a regression representation for these objects and establish
conditions under which they are identified by local
projections with instrumental-variables (IV). As a result, sequential counterfactual outcome paths can be
recovered from a finite collection of local IV regressions.

Second, under a linear envelope and a maintained policy invariance restriction,
we justify the local projection mapping and establish an equivalence result.
Any empirical VAR or forward-looking structural model that implies the same
impulse responses for observables delivers the same sequential policy-path
counterfactuals under our procedure, so conclusions are pinned down by
identified impulse responses rather than by the particular full-system
representation. We justify the invariance restriction by giving a sufficient
economic condition. In a micro-founded model, it holds when the private-sector
block is invariant to policy, so interventions affect private behavior only
through the induced expected path of the policy instrument, see
\cite{mckay2023can}.

Third, we extend the local projection implementation to the one-shot design of
\cite{mckay2023can} and clarify the trade-off across designs. Sequential
counterfactuals are most credible for short windows and moderate interventions,
when expectation revisions are limited. One-shot counterfactuals are
expectations-consistent by construction, but they typically require a richer
set of identified policy shocks to implement the path restriction. The relevant
distinction is the information set of private agents, sequential benchmarks
unanticipated interventions, whereas one-shot benchmarks anticipated policy
paths.

\textit{Relation to the literature.}---This paper relates to recent work that frames policy counterfactuals through a
sufficient-statistic logic. \cite{mckay2023can}, \cite{beraja2023semistructural},
and \cite{barnichon2026policy} study counterfactual policy rules under parametric
restrictions and trace the associated equilibrium implications. Our focus is on
counterfactual policy paths. We provide local-projection identification of the
induced counterfactual trajectories and show that, under policy invariance and
a moving-average envelope, the LP estimands coincide with the counterfactual
objects implied by empirical and structural models that share the same envelope.

The analysis is also connected to VAR-based policy-scenario and
conditional-forecast exercises, including \cite{bernanke1997systematic},
\cite{waggoner1999conditional}, \cite{leeper2003modest}, and
\cite{antolin2021structural}. Recent work shifts attention to identification in
more general, potentially non-invertible, moving-average representations, see
\cite{stock2018identification} and \cite{plagborg2022instrumental}. We formulate
the counterfactual directly at the moving-average level. Under policy
invariance, models that share the same impulse responses deliver identical
counterfactual implications, so the identifying content of the counterfactual
is disciplined by the impulse responses alone.

Our LP identification takes identified structural shocks as instruments, and thus links directly to established strategies for macroeconomic shock identification, including \cite{ramey2016macroeconomic}, \cite{kilian2017structural}, \cite{gertler2015monetary}, \cite{nakamura2018high}, \cite{jarocinski2020deconstructing}, and \cite{bauer2023reassessment}.

\textit{Empirical application.}---We illustrate the framework with two policy counterfactuals. First, we study policy-peg impulse responses for U.S.\ oil-price disturbances under counterfactual short-rate pegs of length three, twelve, and twenty-four months. The oil shock is identified using the oil-supply news surprise of \cite{kanzig2021macroeconomic}, and monetary policy shocks are identified with the high-frequency external instrument of \cite{bauer2023reassessment}. The mapping recovers the shock sequence that implements each peg and shows that the required intervention is largely front-loaded. The exercise quantifies the role of systematic monetary policy in the propagation of oil shocks and delivers a short-run inflation--output trade-off.

Second, we study policy-path effects in the post-pandemic tightening episode through a retrospective liftoff counterfactual. We shift liftoff from February~2022 to November~2021 and vary the intervention length. Impulse responses identified using the high-frequency instrument of \cite{bauer2023reassessment} map an imposed short-rate path deviation into counterfactual paths for real output and inflation, and, as a by-product, deliver the policy-shock sequence that implements the deviation. Earlier and longer liftoff generates larger disinflation alongside a deeper and more persistent contraction in real activity.

\textit{Outline.}---Section \ref{sec2:framework} formalizes policy counterfactuals under a
sequential design. Section \ref{sec3:lp} develops identification via local
projections for the sequential counterfactual objects. Section \ref{sec4:lp}
justifies the LP estimands and establishes their equivalence to the
counterfactual objects delivered by empirical and structural models under a
common linear SVMA envelope. Section \ref{sec:oneshot} extends our LP policy counterfactuals to
one-shot counterfactual designs. Section \ref{sec:miscel} clarifies the scope
and interpretation of the method. Section \ref{sec6:counterfactual monetary
policy} presents two empirical studies on monetary policy counterfactuals and
draws their empirical implications. Section \ref{sec9:concl} concludes.

\section{Sequential policy counterfactuals}
\label{sec2:framework}

This section formalizes the policy counterfactual objects in a stochastic dynamic environment and specifies the intervention design underlying the baseline analysis. The counterfactual is generated by intervening on the policy shock innovation period by period over a finite window, while holding the non policy shocks unaffected. Section~\ref{sec:oneshot} considers a one shot design in which the intervention is concentrated at the event date and the future policy path restriction is internalized in private sector expectations.

Let $\boldsymbol w_t := (x_t,\, y_t,\, z_t,\,\ell_t)'$ denote the state observed
at date $t$. For interpretation, $y_t$ may be output, $x_t$ a variable of
interest, for example an oil price component, $z_t$ a policy instrument, and
$\ell_t$ a block of remaining variables that may be partially unobserved.

We represent outcomes as generated by the sequence of structural shocks through a
measurable mapping
\begin{align}
\boldsymbol w_t = \boldsymbol g(\boldsymbol{\varepsilon}_{-\infty:t}),
\end{align}
where $\boldsymbol{\varepsilon}_{-\infty:t} := (\boldsymbol{\varepsilon}_{s})_{s\le t}$
is the ordered sequence of current and past shocks. Write
$\boldsymbol{\varepsilon}_t := (\varepsilon_{x,t},\,\varepsilon_{z,t},\,\boldsymbol{\varepsilon}_{I,t})$
for the date $t$ shock vector, partitioned into the shock of interest
$\varepsilon_{x,t}$, a policy shock $\varepsilon_{z,t}$, and a vector of
remaining shocks $\boldsymbol{\varepsilon}_{I,t}$. Throughout we maintain
$\boldsymbol{\varepsilon}_t\overset{i.i.d}{\sim}(0,I)$.\footnote{This assumption is imposed
for clarity. It can be relaxed to intertemporal and contemporaneous orthogonality
restrictions stated as moment conditions sufficient for consistent estimation
and valid inference.}

\subsection{Policy-peg impulse responses}

Impulse responses summarize the dynamic effect of a shock of interest in a
stochastic environment. Fix $h\ge 0$ and consider a treatment of the $x$ shock.
The impulse response of $y$ at horizon $h$ is
\footnote{The binary contrast $(1,0)$ is a special case of a continuous
intervention with magnitudes $d+\delta$ and $d$. In nonlinear systems, the
resulting response may depend on both the baseline level $d$ and the
perturbation $\delta$, even after rescaling.}
\begin{align}
\label{irf}
\theta_{y,h}
:=
\mathbb E\!\left[y_{t+h}\mid \varepsilon_{x,t}=1\right]
-
\mathbb E\!\left[y_{t+h}\mid \varepsilon_{x,t}=0\right].
\end{align}
Equation~\eqref{irf} compares two regular conditional expectations that differ
only in the date $t$ realization of $\varepsilon_{x,t}$, integrating out all
other sources of uncertainty. This definition corresponds to the usual
impulse-response experiment and admits a potential-outcome interpretation for
time-series interventions.\footnote{See, among others, \cite{angrist2011causal},
\cite{angrist2018semiparametric},
\cite{bojinov2019time}, and \cite{rambachan2021common}.}

We now define a policy-peg counterfactual that shuts down systematic policy
responses following the $x$ shock. Fix $q\ge 0$, let
$\mathcal Q=\{0,1,\dots,q\}$, and write
$\boldsymbol z_{t,\mathcal Q}:=(z_{t+\ell})_{\ell\in\mathcal Q}$.
The object of interest is the component of the impulse response that operates
through endogenous movements in the policy instrument.

To construct the pegged economy, let $\varepsilon_{z,s}$ denote the policy-shock
innovation and let $\boldsymbol{\varepsilon}_{-z,s}:=(\varepsilon_{x,s},\,\boldsymbol{\varepsilon}_{I,s})$
collect the remaining shocks. For a given $h$ and $\mathcal Q$, define a hypothetical
shock sequence over $\{t,\dots,t+h\}$ by
\begin{align}
\label{cf:shock}
\tilde{\boldsymbol{\varepsilon}}_{t:t+h}
:=
\bigl(
\boldsymbol{\varepsilon}_{-z,t+\ell},
\,
\tilde{\varepsilon}_{z,t+\ell}
+
\mathbf 1_{\{\ell\notin\mathcal Q\}}
\bigl(\varepsilon_{z,t+\ell}-\tilde{\varepsilon}_{z,t+\ell}\bigr)
\bigr)_{0\le \ell\le h}.
\end{align}
Thus, for $\ell\in\mathcal Q$ the policy shock is replaced by its counterfactual
counterpart, while for $\ell\notin\mathcal Q$ it is not. The counterfactual economy is obtained by feeding the counterfactual
shock sequence through the same structural mapping,
\[
\tilde{\boldsymbol w}_{t+h}
=
\boldsymbol g\!\left(\boldsymbol{\varepsilon}_{-\infty:t-1},\,\tilde{\boldsymbol{\varepsilon}}_{t:t+h}\right),
\]
where
$\tilde{\boldsymbol w}_{t+h}:=(\tilde x_{t+h},\,\tilde y_{t+h},\,\tilde z_{t+h},\,\tilde \ell_{t+h})'$. This construction holds $\boldsymbol g$ fixed and alters only the shock sequence.
Economically, $\boldsymbol g$ collects private-sector decision rules together with the baseline policy rule and the information structure that map structural innovations into observables.
Policy invariance therefore keeps equilibrium restrictions unchanged and evaluates outcomes under the same mapping after modifying only the policy shocks within the finite window $\mathcal Q$.

A useful discipline for the sequential design is an innovations interpretation.
Let $\{\mathcal F_t\}$ denote the agents' filtration.
The counterfactual replaces the policy shock at each date $t+\ell$, $\ell\in\mathcal Q$, by a period-by-period surprise that is revealed at $t+\ell$, orthogonal to $\mathcal F_{t+\ell-1}$, and drawn from the same conditional law as in the baseline economy, see detailed discussion in Section \ref{subsec:policy_path_fw_equiv}.
Under this discipline, the policy peg is not a new rule, it is a finite string of temporary deviations from the baseline rule.
The restriction is most strained when the implied vector $\Delta_{\mathcal Q}$ is large, persistent, and one sided, because such a pattern is informative about a regime shift and may trigger belief revision in the sense of \cite{lucas1976econometric}.
This consideration motivates the separate one-shot design in Section~\ref{sec:oneshot}, which allows the imposed path to be internalized in expectations from the outset.

The peg is imposed in conditional expectation through a selection rule indexed
by the date-$t$ realization $d$ of the shock of interest. For each relevant
value $d$, choose a sequence of counterfactual policy shocks
$\{\tilde{\varepsilon}_{z,t+\ell}(d):\ell\in\mathcal Q\}$ so that the expected
policy path in the counterfactual economy over $\mathcal Q$ coincides with the
untreated baseline path,
\begin{align}
\label{peg}
\mathbb E
\!\left[
\tilde{\boldsymbol z}_{t,\mathcal Q}
\mid
\varepsilon_{x,t}=d
\right]
=
\mathbb E
\!\left[
\boldsymbol z_{t,\mathcal Q}
\mid
\varepsilon_{x,t}=0
\right].
\end{align}
We normalize the untreated case by setting
$\tilde{\varepsilon}_{z,t+\ell}(0)=\varepsilon_{z,t+\ell}$ for all
$\ell\in\mathcal Q$, so the pegged economy coincides with the baseline economy
under $\varepsilon_{x,t}=0$. This definition restricts attention to feasible counterfactual designs: for the horizons under consideration and the relevant values of $d$, there exists an admissible within-window policy-shock interventions that implements the targeted restriction on the policy-instrument path, holding the non-policy shocks fixed. Section~\ref{sec4:lp} provides a structural feasibility condition and shows that, under the linear SVMA envelope, the implementing intervention sequence is uniquely pinned down.

Finally, the policy-peg impulse response at horizon $h$ is defined as
\begin{align}
\label{irf-peg}
\beta_{h,\mathcal Q}
:=
\mathbb E
\!\left[ \tilde y_{t+h} \mid \varepsilon_{x,t}=1 \right]
-
\mathbb E
\!\left[ \tilde y_{t+h} \mid \varepsilon_{x,t}=0 \right].
\end{align}
Conditioning on $\varepsilon_{x,t}$ is interpreted as the impulse-response
experiment applied within the pegged economy. The object $\beta_{h,\mathcal Q}$
therefore compares treated and untreated potential outcomes under the same
policy restriction and isolates the dynamic effect of the $x$ shock when the
expected policy path over $\mathcal Q$ is fixed at its untreated evolution.

\subsection{Policy-path effects}

This subsection defines a policy-path effect in a realization-based form.
Fix an evaluation horizon $h\ge 0$ and condition on a realized sequence of
structural shocks. The construction contrasts the realized outcome
generated by the observed shock sequence with a counterfactual outcome obtained
by replacing the policy-shock realizations over a finite window, so as to
deliver a prescribed policy-instrument path.

Let $\boldsymbol e_{-\infty:t+h}:=(\boldsymbol e_s)_{s\le t+h}$ denote a realized
sequence of the structural-shock vector, where $\boldsymbol e_s$ is a realization
of $\boldsymbol{\varepsilon}_s$. The baseline outcome at horizon $h$ is the
realized value generated by this shock sequence,\footnote{Equivalently,
$\boldsymbol w_{t+h}^{path}$ can be written as a conditional expectation
given the realized shock sequence, since the conditional distribution degenerates
at the realized value.}
\begin{align}
\label{eq:baseline-outcome}
\boldsymbol w_{t+h}^{path}
:=
\boldsymbol g\!\left(\boldsymbol e_{-\infty:t+h}\right).
\end{align}
When the object of interest is $y_{t+h}$, write $y_{t+h}^{path}$ for the
corresponding component of $\boldsymbol w_{t+h}^{path}$.

Let
$\boldsymbol z_{t,\mathcal Q}^{\mathrm{base}}:=(z_{t+\ell})_{\ell\in\mathcal Q}$
be the realized baseline policy path implied by the realized shock history
$\boldsymbol e_{-\infty:t+h}$. The scenario is indexed by a prescribed path gap,
or deviation, vector $\boldsymbol c_{\mathcal Q}:=(c_\ell)_{\ell\in\mathcal Q}$,
defined as the difference between the counterfactual and baseline paths over $\mathcal Q$, and targets
\begin{align}
\label{eq:peg-constraint}
\tilde{\boldsymbol z}_{t,\mathcal Q}^{path}(\boldsymbol c_{\mathcal Q})
=
\boldsymbol z_{t,\mathcal Q}^{path}
+
\boldsymbol c_{\mathcal Q}.
\end{align}
The restriction \eqref{eq:peg-constraint} is imposed only over $\mathcal Q$ and
is implemented by replacing the realized policy-shock sequence on $\mathcal Q$,
in the same shock-sequence replacement sense as in \eqref{cf:shock}. The
experiment therefore alters only the realized policy-shock component over
$\mathcal Q$ while leaving the realized values of all remaining shocks
unchanged.

Formally, let $e_{z,s}$ denote the realized policy shock at date $s$, and let
$\boldsymbol e_{-z,s}$ collect the realizations of all remaining shocks. Fix
$h\ge 0$ and define the counterfactual realized shock sequence
$\tilde{\boldsymbol e}_{t:t+h}:=(\tilde{\boldsymbol e}_{t+\ell})_{\ell=0}^h$ by
imposing $\tilde{\boldsymbol e}_{-z,t+\ell}=\boldsymbol e_{-z,t+\ell}$ for all
$\ell=0,1,\dots,h$, and $\tilde e_{z,t+\ell}=e_{z,t+\ell}$ for all
$\ell\notin\mathcal Q$, while for $\ell\in\mathcal Q$ we replace $e_{z,t+\ell}$
by a counterfactual value $\tilde e_{z,t+\ell}$. The counterfactual values
$\{\tilde e_{z,t+\ell}:\ell\in\mathcal Q\}$ are chosen so that the induced
policy-instrument path satisfies \eqref{eq:peg-constraint}. This presumes that
the targeted path is feasible under the maintained mapping $\boldsymbol g$, and
when the implementing policy-shock sequence is not unique, the definition
requires an admissible selection rule.

The hypothetical outcome at horizon $h$ is obtained by feeding the counterfactual
shock sequence through the same law of motion. Writing $g_y$ for the $y$
coordinate of $\boldsymbol g$, define
$\tilde y_{t+h}^{path}(\boldsymbol c_{\mathcal Q})
:=
g_y\!\left(\boldsymbol e_{-\infty:t-1},\,\tilde{\boldsymbol e}_{t:t+h}\right),$
so that the counterfactual differs from the observed realization only through
the policy-shock component over the constrained horizons. The policy-path effect
at horizon $h$ is then summarized by
\begin{align}
\label{eq:policy-path-effect}
\delta_h(\boldsymbol c_{\mathcal Q})
:=
\tilde y_{t+h}^{path}(\boldsymbol c_{\mathcal Q})
-
y_{t+h}^{path}.
\end{align}

The distinction from policy-peg impulse responses has two dimensions, the experimental design and the conditioning target. Policy-peg impulse responses are conditional expectations under an impulse-response experiment that assigns a one unit innovation in the shock of interest at the event date, and then imposes the policy restriction in expectation. Policy-path effects are intrinsically pathwise, they dispense with any impulse-response experiment and instead evaluate counterfactual outcomes conditional on a realized shock history, imposing the restriction on the realized instrument path. Both objects rest on the same maintained policy-invariance restriction, the same structural mapping, and the same shock-sequence replacement device.

\section{Local projection identification}
\label{sec3:lp}

This section introduces local projection estimands that identify sequential
policy counterfactuals. The intervention is a restriction on the policy
instrument over a finite window, so the restricted object is the post-treatment
stack $\boldsymbol z_{t,\mathcal Q}$. For this reason, identification cannot be
phrased as exogeneity of the policy path. Instead, we use within-window policy
shocks as excluded variation and impose stacked IV moments that pin down the
coefficients on $\boldsymbol z_{t,\mathcal Q}$ and on the shock of interest.

\subsection{Policy-peg impulse responses}
Given the intervention window $\mathcal Q$, collect the within-window policy path as $\boldsymbol z_{t,\mathcal Q}:=(z_t,\dots,z_{t+q})'$. For each horizon $h\ge 0$, we define the LP-IV estimand through the population local projection
\begin{align}
\label{eq:lp-peg}
y_{t+h}
=
\beta_{h,\mathcal Q}^{LP}\,\varepsilon_{x,t}
+
(\boldsymbol\gamma_{h,\mathcal Q}^{LP})'\boldsymbol z_{t,\mathcal Q}
+
u_{t,h,\mathcal Q},
\end{align}
where $u_{t,h,\mathcal Q}$ is the population IV residual associated with the coefficients defined by the moment restrictions below. The regression is not an equilibrium restriction and not a forecasting device. The role of $\boldsymbol z_{t,\mathcal Q}$ is purely implementational, it encodes the peg restriction and isolates the endogenous within-window movement that the counterfactual shuts down.

Identification uses the sequence of policy shocks as instruments to isolate the excluded component of policy variation. Let
$\boldsymbol{\varepsilon}_{z,t,\mathcal Q}:=(\varepsilon_{z,t},\dots,\varepsilon_{z,t+q})'$
denote the within-window policy-shock sequence and define the instrument stack
$\mathcal E_{t,\mathcal Q}:=(\varepsilon_{x,t},\boldsymbol{\varepsilon}_{z,t,\mathcal Q}')'$.
We maintain the IV exogeneity restriction
\begin{align}
\mathbb E\!\left[\mathcal E_{t,\mathcal Q}\left(y_{t+h}
-
\beta_{h,\mathcal Q}^{LP}\varepsilon_{x,t}
-
(\boldsymbol\gamma_{h,\mathcal Q}^{LP})'\boldsymbol z_{t,\mathcal Q}
\right)\right]=0.
\end{align}
This moment system defines $\beta_{h,\mathcal Q}^{LP}$ and $\boldsymbol\gamma_{h,\mathcal Q}^{LP}$ as population IV coefficients and, through \eqref{eq:lp-peg}, defines $u_{t,h,\mathcal Q}$ as the associated residual. Section~\ref{sec4:lp} provides a linear SVMA envelope under which the maintained unconditional moments can be justified from independence of the structural shocks.

Let $X_{t,\mathcal Q}:=(\varepsilon_{x,t},\boldsymbol z_{t,\mathcal Q}')'$ denote
the regressor stack in \eqref{eq:lp-peg} and define
$\Bar\Sigma_{\mathcal Q}:=\mathbb E[\mathcal E_{t,\mathcal Q}X_{t,\mathcal Q}']$.
In particular, the within-window policy shocks satisfy
$\mathbb E[\boldsymbol{\varepsilon}_{z,t,\mathcal Q}u_{t,h,\mathcal Q}]=0$.
To discipline the rank requirement for the post-treatment stack, we appeal to
the within-window period-by-period intervention design in Assumption~\ref{ass:structural-env} (iii): the date-$t+\ell$
policy shock shifts $z_{t+\ell}$ on impact and leaves $\{z_t,\dots,z_{t+\ell-1}\}$
unaffected. With this timing, any nonzero contemporaneous shock effect on the
policy instrument delivers a generic nonsingularity condition for
$\Bar\Sigma_{\mathcal Q}$.

Under these maintained restrictions, let $\lambda=(1,0,\ldots,0)'$ select the
coefficient on the shock of interest. The policy-peg response is identified as
\begin{align}
\label{eq:beta-id}
\beta_{h,\mathcal Q}^{LP}
=
\lambda'\bar \Sigma_{\mathcal Q}^{-1}\bar \Gamma_{h,\mathcal Q},
\end{align}
where $\bar \Gamma_{h,\mathcal Q}:=\mathbb E[\mathcal E_{t,\mathcal Q}y_{t+h}]$.
Equation \eqref{eq:beta-id} makes the identifying variation explicit. The peg is
encoded through inclusion of $\boldsymbol z_{t,\mathcal Q}$, while excluded
within-window policy shocks identify the component of the projection that is
orthogonal to endogenous policy variables over $\mathcal Q$.

The interpretation links to the sequential counterfactual in
Section~\ref{sec2:framework}. As formalized in \eqref{cf:shock}, the peg
experiment introduces within-window policy shocks over $\mathcal Q$ that enforce
the peg restriction in \eqref{peg}. Under linearity, implementation corresponds
to offsets that undo the movement in $\boldsymbol z_{t,\mathcal Q}$ induced by
$\varepsilon_{x,t}$. In \eqref{eq:lp-peg}, controlling for the within-window
policy-path stack $\boldsymbol z_{t,\mathcal Q}$ removes from the
$\varepsilon_{x,t}$ coefficient the within-window policy-channel component of
the $x$-shock response.

The same identification logic applies when the within-window policy shocks are latent and the researcher observes a proxy $s_t$.
Let $\boldsymbol s_{t,\mathcal Q}:=(s_t,\ldots,s_{t+q})'$ and use it in place of $\mathcal E_{t,\mathcal Q}$ as the excluded instrument stack.
In addition to the exclusion restriction, identification requires a relevance condition for the stacked problem, $\operatorname{rank}\,\mathbb E[\boldsymbol s_{t,\mathcal Q}\boldsymbol z_{t,\mathcal Q}']=q+1$.
A sufficient primitive condition is $s_t=\pi\,\varepsilon_{z,t}+\nu_t$ with $\pi\neq 0$ and $\nu_t$ orthogonal to the structural shocks, under the maintained timing restrictions this scalar relevance lifts to the full regressor stack.

\subsection{Policy-path responses}

The policy-peg coefficient $\beta_{h,\mathcal Q}^{LP}$ captures the response to
the shock of interest, holding fixed the realized policy-instrument stack in
\eqref{eq:lp-peg}. A complementary object is the policy-path response
coefficients $\boldsymbol\gamma_{h,\mathcal Q}^{LP}$, which map a within-window
policy-instrument deviation into its horizon-$h$ outcome effect. For an
admissible path deviation $\boldsymbol c_{\mathcal Q}$, define the associated LP estimand of policy-path effect
\begin{align}
\label{eq:path-def}
\delta_h^{LP}(\boldsymbol c_{\mathcal Q})
=
(\boldsymbol\gamma_{h,\mathcal Q}^{LP})'\boldsymbol c_{\mathcal Q}.
\end{align}
Here, $\delta_h^{LP}(\boldsymbol c_{\mathcal Q})$ is the induced horizon-$h$
effect under the path deviation $\boldsymbol c_{\mathcal Q}$.

To identify $\boldsymbol\gamma_{h,\mathcal Q}^{LP}$, we rewrite \eqref{eq:lp-peg}
by absorbing the shock-of-interest term into the regression error. Let
$v_{t,h,\mathcal Q}:=u_{t,h,\mathcal Q}+\beta_{h,\mathcal Q}^{LP}\varepsilon_{x,t}$.
Then \eqref{eq:lp-peg} implies
\begin{align}
\label{eq:lp-path-reg}
y_{t+h}
=
(\boldsymbol\gamma_{h,\mathcal Q}^{LP})'\boldsymbol z_{t,\mathcal Q}
+
v_{t,h,\mathcal Q}.
\end{align}
This reparameterization is purely algebraic and adds no identifying content.
Identification of $\boldsymbol\gamma_{h,\mathcal Q}^{LP}$ uses the same excluded
policy-shock variation $\boldsymbol\varepsilon_{z,t,\mathcal Q}$ as
\eqref{eq:lp-peg}, it requires no additional instruments and no additional
exclusion restrictions. Under the maintained moment conditions
$\mathbb E[\boldsymbol\varepsilon_{z,t,\mathcal Q}\,u_{t,h,\mathcal Q}]=0$ and
$\mathbb E[\boldsymbol\varepsilon_{z,t,\mathcal Q}\,\varepsilon_{x,t}]=0,$ the
definition of $v_{t,h,\mathcal Q}$ yields
$\mathbb E[\boldsymbol\varepsilon_{z,t,\mathcal Q}\,v_{t,h,\mathcal Q}]=0$.
Together with nonsingularity of
$\Sigma_{\mathcal Q}:=\mathbb E[\boldsymbol\varepsilon_{z,t,\mathcal Q}\boldsymbol z_{t,\mathcal Q}']$,
the policy-path responses satisfy
\begin{align}
\label{eq:gamma-id}
\boldsymbol\gamma_{h,\mathcal Q}^{LP}
=
\Sigma_{\mathcal Q}^{-1}\Gamma_{h,\mathcal Q},
\end{align}
where $\Gamma_{h,\mathcal Q}:=\mathbb E[\boldsymbol\varepsilon_{z,t,\mathcal Q}\,y_{t+h}].$

This representation clarifies the sequential design. The policy-path responses
$\boldsymbol\gamma_{h,\mathcal Q}^{LP}$ summarize how within-window movements in
the policy instrument map into $y_{t+h}$. Under the maintained timing
restriction, each date-$t+\ell$ policy shock shifts $z_{t+\ell}$ on impact and
does not affect earlier policy realizations, so $\Sigma_{\mathcal Q}$ is
generically nonsingular and pins down $\boldsymbol\gamma_{h,\mathcal Q}^{LP}$
through \eqref{eq:gamma-id}.

Thus far, $\beta_{h,\mathcal Q}^{LP}$ and $\boldsymbol\gamma_{h,\mathcal Q}^{LP}$
are defined and identified as population IV projection coefficients under
\eqref{eq:lp-peg} and the maintained moment conditions under the sequential
design. Section~\ref{sec4:lp} provides a linear moving average
envelope under which these LP estimands coincide with the corresponding
sequential counterfactual objects defined in Section~\ref{sec2:framework}.

\section{Structural interpretation of LP counterfactuals}
\label{sec4:lp}

Section~\ref{sec3:lp} identifies the LP estimands that enter our counterfactual
constructions. This section links these projection coefficients to the
structural experiments in Section~\ref{sec2:framework} and states maintained
restrictions under which the LP estimands coincide with the corresponding
counterfactual objects.

We work within a linear SVMA envelope with time-invariant impulse
responses, impose policy invariance across the baseline and counterfactual
economies, and consider finite-horizon interventions in the policy-shock
sequence. Under these restrictions, LP and structural counterfactuals agree.

The argument proceeds in three steps. We introduce the envelope, characterize
finite-horizon interventions and show that, under the maintained restrictions,
the LP estimands coincide with the corresponding counterfactual objects and
admit an algebraic representation in terms of impulse responses, and discuss
two standard implementations, an empirical SVAR recursion and a forward-looking
structural model, that both fall under the envelope.

\subsection{An SVMA envelope}
\label{subsec:svma-envelope}

We embed the structural mapping
$\boldsymbol w_t=\boldsymbol g(\boldsymbol\varepsilon_{-\infty:t})$
in a linear SVMA envelope,
\begin{align}
\label{eq:svma}
\boldsymbol w_t
=
\sum_{j=0}^{\infty}\Theta_j\,\boldsymbol\varepsilon_{t-j},
\end{align}
where $\Theta_0$ is of full row rank and the coefficient matrices are absolutely
summable in spectral norm, $\sum_{j=0}^{\infty}\|\Theta_j\|_2<\infty$. With
$\mathbb E[\boldsymbol\varepsilon_t]=0$ and
$\mathbb E[\boldsymbol\varepsilon_t\boldsymbol\varepsilon_t']$ is diagonal,
$\{\boldsymbol w_t\}$ is a causal second-order stationary linear process. 
The role of \eqref{eq:svma} is not to impose a particular identification scheme.
It is a maintained envelope that imposes linearity and delivers time-invariant
impulse responses, which summarize the dynamic response of observables to
the shock of interest and to the policy shock.

Recoverability is a separate issue. When the dimension of
$\boldsymbol\varepsilon_t$ equals the dimension of $\boldsymbol w_t$ and the
moving-average lag polynomial is invertible, \eqref{eq:svma} admits an
equivalent SVAR representation, so shocks are recoverable from VAR innovations,
subject to the usual rotational indeterminacy under an admissible
identification. When the number of shocks exceeds the number of observables,
the system is a nonrecoverable SVMA in the sense of \cite{stock2018identification},
and shocks are not uniquely pinned down by observables alone. Our counterfactual
results do not require recoverability. They condition on the relevant impulse
responses as equilibrium objects delivered by an admissible identification
strategy, including VAR-based identification, proxy-SVAR procedures, or direct
LP--IV designs. Conditional on these inputs, the counterfactual mapping is a
deterministic implication of linearity and the maintained intervention class.

The envelope \eqref{eq:svma} covers the linear structures routinely used in
macroeconomics. A stable state-space system with spectral radius strictly inside the unit circle
admits an SVMA representation in structural shocks, see Chapter~2 of
\cite{ljungqvist2018recursive}. Likewise, a first-order approximation of a
rational-expectations model yields a stable solution under the
Blanchard--Kahn determinacy conditions, and this solution can be written as an
SVMA in structural shocks, see Chapter~11 of \cite{ljungqvist2018recursive} and
\cite{blanchard1980solution}.  If the solution is invertible, it also admits a
Wold moving-average representation and an associated VAR
representation.

\subsubsection{From policy shocks to policy paths}

We formalize counterfactual policy experiments as finite-horizon interventions
in the policy-shock sequence. The maintained object is the equilibrium mapping
that propagates shocks into outcomes. The counterfactual therefore differs from
the baseline only through a modified policy-shock sequence over an intervention
window $\mathcal Q$.

\begin{assumption}[Structural counterfactual environment]
\label{ass:structural-env}\leavevmode
\begin{enumerate}[(i)]
\item \textup{(Linear representation.)}
The economy admits a linear moving-average representation of the form
\eqref{eq:svma}.
\item \textup{(Policy invariance.)}
The structural mapping $\boldsymbol g$ is invariant across the baseline and
counterfactual economies.
\item \textup{(Sequential shock intervention.)} For each $\ell\in\mathcal Q$, the intervention operates only through the policy shock according to $\tilde{\varepsilon}_{z,t+\ell}=\varepsilon_{z,t+\ell}+\delta_\ell$, for all $\ell\in \mathcal{Q}$ and leaves all non-policy shocks unchanged.
\end{enumerate}
\end{assumption}

Assumption~\ref{ass:structural-env} (\textit{ii}) formalizes policy invariance.
The counterfactual keeps the equilibrium mapping $\boldsymbol g$ fixed and modifies only the policy-shock sequence over a finite window $\mathcal Q$.
Assumption~\ref{ass:structural-env} (\textit{iii}) makes the restriction operational by perturbing only the policy-shock component.
The intervention is finite-horizon in the sense that the shock modification is confined to $\mathcal Q$, while induced deviations in endogenous variables may propagate beyond $\mathcal Q$ through the system dynamics. In forward-looking environments, the sequential design admits a period-by-period surprise interpretation relative to the baseline filtration.

This assumption has a transparent intervention interpretation. The policy-path counterfactual is implemented by a finite sequence of policy-shock interventions that reproduces the desired path under the maintained baseline law of motion. This formulation is in the spirit of the intervention analysis in \cite{box1975intervention}. Related counterfactual analyses adopt the similar logic. For instance, \cite{mckay2023can} show that a rule-change counterfactual can be represented by an appropriate sequence of policy shocks evaluated under the original mapping, while the counterfactual policy variable itself satisfies the equations of the alternative rule.

For the intervention window $\mathcal Q$ introduced above, let $\Delta_{\mathcal Q}:=(\delta_\ell)_{\ell\in\mathcal Q}$ denote the sequence of policy interventions in Assumption~\ref{ass:structural-env}(\textit{iii}). Combining Assumption~\ref{ass:structural-env}(\textit{i}) with the intervention in (\textit{iii}) implies the policy-path restriction,
\begin{align}
\label{eq:path-restriction}
\Psi(\mathcal Q)\,\Delta_{\mathcal Q}+\boldsymbol z_{t,\mathcal Q}
=
\tilde{\boldsymbol z}_{t,\mathcal Q}.
\end{align}
where $\Psi(\mathcal Q)$ is the $(q+1)\times(q+1)$ lower-triangular Toeplitz
matrix generated by the impulse responses
$\psi_{z,h}:=\partial z_{t+h}/\partial \varepsilon_{z,t}$, with $\psi_{z,h}=0$
for $h<0$. Equivalently,
$\Psi(\mathcal Q)=[\psi_{z,i-j}]_{i,j=1,\ldots,q+1}$.
Since $\Psi(\mathcal Q)$ is lower triangular with diagonal entries $\psi_{z,0}$,
implementability and uniqueness of the within-window shock intervention are
equivalent to $\psi_{z,0}\neq 0$, or, equivalently, to nonsingularity of
$\Psi(\mathcal Q)$. This restriction is the structural counterpart of the rank condition in the LP--IV formulation and it formalizes the existence and uniqueness of the sequential counterfactual construction in Section~\ref{sec2:framework}.

\subsubsection{Counterfactual characterization}

\textit{Policy-peg impulse response.}\\
The peg restriction \eqref{peg} requires that, under the $x$-shock treatment,
the counterfactual policy path over $\mathcal Q$ coincides with the baseline
path that would obtain absent the treatment, that is,
$\mathbb{E}\!\left[\tilde{\boldsymbol z}_{t,\mathcal{Q}}\mid \varepsilon_{x,t}=1\right]
=
\mathbb{E}\!\left[{\boldsymbol z}_{t,\mathcal{Q}}\mid \varepsilon_{x,t}=0\right]$.
Under Assumption~\ref{ass:structural-env} and linearity, a stacked intervention
${\Delta}_{\mathcal Q}^{peg}$ shifts the conditional mean of the policy path by
$\Psi(\mathcal Q){\Delta}_{\mathcal Q}^{peg}$. Conditioning on
$\varepsilon_{x,t}=1$ yields
\begin{align}
\label{eq:mean-shift-mq-general}
\Psi(\mathcal{Q}){\Delta}_{\mathcal Q}^{peg}
+
\mathbb{E}\!\left[{\boldsymbol z}_{t,\mathcal{Q}}\mid \varepsilon_{x,t}=1\right]
=
\mathbb{E}\!\left[\tilde{\boldsymbol z}_{t,\mathcal{Q}}\mid \varepsilon_{x,t}=1\right].
\end{align}
Imposing the peg restriction and rearranging,
\begin{align}
\label{eq:mean-shift-mq}
\Psi(\mathcal{Q}){\Delta}_{\mathcal Q}^{peg}
=-\boldsymbol\theta_{z}(\mathcal{Q}),
\end{align}
where $\boldsymbol\theta_{z}(\mathcal Q):=
\mathbb{E}\!\left[{\boldsymbol z}_{t,\mathcal{Q}}\mid \varepsilon_{x,t}=1\right]
-
\mathbb{E}\!\left[{\boldsymbol z}_{t,\mathcal{Q}}\mid \varepsilon_{x,t}=0\right]
=(\theta_{z,0},\ldots,\theta_{z,q})'$ stacks the impulse responses
of the policy instrument along $\mathcal Q$, as defined in \eqref{eq:path-restriction}. As $\Psi(\mathcal Q)$ is
nonsingular, \eqref{eq:mean-shift-mq} pins down ${\Delta}_{\mathcal Q}^{peg}$
uniquely.

Given ${\Delta}_{\mathcal Q}^{\mathrm{peg}}$, the associated counterfactual response of the outcome at horizon $h$ can be interpreted as the effect of the initial $x$-treatment, together with the sequence of policy interventions ${\Delta}_{\mathcal Q}^{\mathrm{peg}}$ that pegs the policy instrument at its pre-treatment level over the window $\mathcal Q$. In a linear environment, the counterfactual response inherits an additivity property, so that the overall effect decomposes into the direct response to the initial treatment and the cumulative contribution of the induced policy interventions,
\begin{align}
\label{cir:post}
\beta_{h,\mathcal Q}
=
\theta_{y,h}
+\boldsymbol{\psi}_{y,h,\mathcal{Q}}^\prime {\Delta}_{\mathcal Q}^{peg},
\end{align}
where $\theta_{y,h}=\partial y_{t+h}/\partial \varepsilon_{x,t}$ is the impulse
response to the period-$t$ shock of interest,
$\boldsymbol{\psi}_{y,h,\mathcal Q}=(\psi_{y,h},\cdots,\psi_{y,h-q})'$ collects the responses that transmit policy-shock interventions over $\mathcal Q$
into $y_{t+h}$, and $\psi_{y,h}=\partial y_{t+h}/\partial \varepsilon_{z,t}$.\\

\noindent\textit{Policy-path effects.}\\
The policy-path effect measures how the outcome would have evolved had the
policy instrument followed a prescribed path over an intervention window. We
interpret this deviation as arising from replacing the realized policy shocks
within the window. The experiment evaluates the same structural mapping
$\boldsymbol g$ along two shock histories that coincide outside the window, a
baseline history ${\boldsymbol e_{t+\ell}}$ and a counterfactual history
${\tilde{\boldsymbol e}_{t+\ell}}$, where only the policy-shock component
differs for $\ell\in\mathcal Q$.

Recall that $\boldsymbol c_{\mathcal Q}$ denotes the prescribed path deviation of the
policy instrument over $\mathcal Q$, stacked conformably with
$\boldsymbol z_{t,\mathcal Q}$. Under \eqref{eq:path-restriction}, the
restriction translates into
\begin{align}
\label{eq:m-diff}
\Psi(\mathcal{Q})\,
\Delta_{\mathcal{Q}}^{path}=\boldsymbol c_{\mathcal Q},
\end{align}
and hence $\Delta_{\mathcal{Q}}^{path}=\Psi(\mathcal{Q})^{-1}\boldsymbol c_{\mathcal Q}$
whenever $\Psi(\mathcal Q)$ is nonsingular. Mapping these implied shock
interventions into the horizon-$h$ outcome yields the implied policy-path effect
\begin{align}
\label{eq:delta-realization}
\delta_h(\boldsymbol c_{\mathcal Q})
:=
\tilde y_{t+h}^{path}(\boldsymbol c_{\mathcal Q})
-
y_{t+h}^{path}=
\boldsymbol{\psi}_{y,h,\mathcal{Q}}'
\Delta_{\mathcal{Q}}^{path}.
\end{align}
Combining \eqref{eq:m-diff} and \eqref{eq:delta-realization} shows that
$\delta_h(\boldsymbol c_{\mathcal Q})$ is linear in $\boldsymbol c_{\mathcal Q}$.
The policy-path response is
$\boldsymbol\gamma_{h,\mathcal Q}:=\Psi(\mathcal Q)^{-1\prime}\boldsymbol{\psi}_{y,h,\mathcal Q}$,
so that $\delta_h(\boldsymbol c_{\mathcal Q})=\boldsymbol\gamma_{h,\mathcal Q}'\boldsymbol c_{\mathcal Q}$.

Equation \eqref{eq:delta-realization} makes explicit what the policy-path
restriction entails under Assumption~\ref{ass:structural-env}. Imposing an
alternative policy path over $\mathcal Q$ is equivalent to selecting a sequence
of policy interventions $\Delta_{\mathcal{Q}}^{path}$ that rationalizes the
restriction through the policy response map $\Psi(\mathcal Q)$. Because the
intervention operates solely through the policy-shock component, the implied
policy-path effect depends only on the impulse responses of the outcome to the
policy shock, summarized by $\boldsymbol{\psi}_{y,h,\mathcal{Q}}$, together with
the within-window policy responses collected in $\Psi(\mathcal Q)$.

The next theorem records the identification statement. Under
Assumption~\ref{ass:structural-env}, the LP estimands target the same
counterfactual objects characterized above.

\begin{theorem}[Local projection identification of the sequential policy counterfactuals]
\label{expostprop}
Suppose Assumption~\ref{ass:structural-env} holds.
Then $\beta_{h,\mathcal Q}=\beta_{h,\mathcal Q}^{LP}$ and
$\boldsymbol\gamma_{h,\mathcal Q}= \boldsymbol\gamma_{h,\mathcal Q}^{LP}$.
\end{theorem}
See the proof in Appendix \ref{proof:expost}. Within the SVMA envelope, policy counterfactuals reduce to a finite collection
of impulse responses. The within-window path restriction pins down the required
sequence of policy interventions, and the counterfactual outcome then follows
by superposition. Concretely, given $\{\theta_{y,h},\theta_{z,h}\}$ for the shock
of interest and $\{\psi_{y,h},\psi_{z,h}\}$ for the policy shock, the identifying
content resides in these impulse responses together with the maintained
restriction that the intervention is implementable through policy innovations
while the equilibrium mapping is held fixed.

This sufficient-statistics perspective delineates the scope of the approach. Outside the SVMA envelope,
nonlinearities and state dependence generally invalidate additivity and can
render impulse responses state dependent. In that case, the mapping from a
targeted path restriction to outcomes no longer reduces to a superposition over
a fixed set of impulse responses, and implementability through a sequence of
policy innovations is no longer guaranteed.

\subsection{Empirical SVAR-based counterfactuals}

In the applied literature, counterfactual exercises are often implemented within a structural VAR for observables,
\begin{align}
\label{eq:svar-emp}
A(L)\boldsymbol w_t=\boldsymbol\varepsilon_t,
\end{align}
where $A(L)$ is stable and invertible, with inverse moving-average operator
$A(L)^{-1}=\Theta(L)$. Under this maintained specification, the SVAR is equivalent to the SVMA envelope \eqref{eq:svma}. A standard SVAR counterfactual selects a
sequence of policy shocks that enforces a within-window restriction on the
policy instrument, while holding fixed the equilibrium mapping, see
\cite{sims1995does}. The difference between the policy-peg and policy-path
exercises is entirely in the within-window restriction.

A necessary requirement for comparing SVAR-based and LP-based counterfactuals
is that the policy shock be the same object across implementations. We take the
impulse responses as equilibrium inputs and normalize them to a common unit
policy shock. Accordingly, the SVAR recursion and the LP--IV construction are
comparable only insofar as they use the same impulse-response functions for the
policy shock. Identification schemes that imply different impulse responses
correspond to different shocks and therefore induce different counterfactual
mappings.

Under the SVAR, the within-window restriction is equivalent to the triangular
system
\begin{equation}
\label{eq:recursive-alg}
\renewcommand{\arraystretch}{1}
\begin{array}{l r c l}
\multirow{1}{*}{\text{period $0$}\quad \quad}
& \tilde{z}_{t}
& = & {z}_{t}+\psi_{z,0}\delta_{0}  \\
\multirow{1}{*}{\text{period $1$}}
& \tilde{z}_{t+1}
& = & {z}_{t+1}
      +\psi_{z,1}\delta_{0}+\psi_{z,0}\delta_{1} \\
& & \vdots & \\
\multirow{1}{*}{\text{period $q$}}
& \tilde{z}_{t+q}
& = & {z}_{t+q}
      +\sum_{\ell=0}^{q}\psi_{z,\ell}\,\delta_{q-\ell}
\end{array}
\end{equation}
which characterizes $\Delta_{\mathcal Q}$ in terms of the policy-shock impulse
responses $\{\psi_{z,\ell}\}_{\ell=0}^q$. Provided $\psi_{z,0}\neq 0$,
\eqref{eq:recursive-alg} admits a unique recursive solution for
$\{\delta_\ell\}_{\ell=0}^q$. This condition coincides with nonsingularity of
$\Psi(\mathcal Q)$ in \eqref{eq:path-restriction}, since \eqref{eq:recursive-alg}
is the lower-triangular representation of the stacked Toeplitz system.

Once the implied within-window intervention sequence $\Delta_{\mathcal Q}$ is
determined, the counterfactual path follows by superposition. The counterfactual
economy equals the baseline economy plus the dynamic effects of the compensating
policy innovations required to satisfy the within-window restriction. Each
innovation enters with its calendar-time alignment through the impulse responses
of the outcome to the policy shock of interest. The response of the policy
instrument to the policy shock pins down $\Delta_{\mathcal Q}$ under the
within-window restriction, and the response of the outcome to the same shock
maps this sequence into the counterfactual outcome path.

The SVAR recursion \eqref{eq:recursive-alg} and the LP construction in
Section~\ref{sec3:lp} are two representations of the same counterfactual mapping.
Under a common SVMA envelope, the counterfactual depends only on linearity,
the within-window restriction, and the associated impulse responses.
LP--IV, which conditions on the within-window policy-path
stack over $\mathcal Q$ and instruments it with policy-shock variation, targets
in population the same mapping that SVAR implementations compute by recursion.

This equivalence has direct implications for practice. First, explicit recursion
is optional, one may compute counterfactuals either by SVAR recursion or by
evaluating the explicit formula induced by the LP regression. The LP coefficients
admit a transparent economic interpretation as mapping coefficients, they map
within-window policy-path deviations into counterfactual outcomes under the
linear envelope. Second, the regression representation records the moments
and instruments that identify each component of the mapping, which facilitates
robustness checks across identification schemes and samples.

\subsection{A forward-looking structural counterfactuals}
\label{subsec:policy_path_fw_equiv}

The counterfactuals can also be constructed from a linearized forward-looking structural
model, such as a rational-expectations system defined with respect to an
increasing filtration $\{\mathcal F_t\}_{t\in\mathbb Z}$, e.g., see \cite{laseena2011anticipated} and \cite{del2013dsge}. At date $t$, the
equilibrium conditions admit the stacked representation
\begin{align}
\label{eq:local-stacked}
\mathcal B_w\,\boldsymbol w^{\,t} + \boldsymbol\varepsilon_{t} = 0,
\end{align}
where $\boldsymbol w^{\,t}
:=
\bigl(\boldsymbol w_{t},\ \mathbb E_{t}\boldsymbol w_{t+1},\ 
\mathbb E_{t}\boldsymbol w_{t+2},\ \ldots \bigr)'$,
$\mathbb E_{t}[\cdot]=\mathbb E[\cdot\mid\mathcal F_{t}]$, and
$\boldsymbol\varepsilon_t$ denotes the date-$t$ structural innovation.
We maintain the standard timing convention
$\mathcal F_t=\sigma(\mathcal F_{t-1},\boldsymbol\varepsilon_t)$ with
$\mathbb E[\boldsymbol\varepsilon_t\mid\mathcal F_{t-1}]=0$, so that the
innovation is the sole source of forecast revision between $t-1$ and $t$.

The impulse-response matrices $\{\Theta_h\}_{h\ge 0}$ admit the
forecast-revision characterization
$\mathbb E_t \boldsymbol w_{t+h}
-
\mathbb E_{t-1} \boldsymbol w_{t+h}
=
\Theta_h\,\boldsymbol\varepsilon_t$.
These matrices summarize equilibrium propagation through contemporaneous
effects and expectational feedback. When the equilibrium admits an innovations
moving-average representation \eqref{eq:svma}, the coefficients in that
representation coincide with $\{\Theta_h\}_{h\ge 0}$. Hence the forward-looking
system and the SVAR recursion are alternative representations of a common
moving-average envelope and therefore supply the same impulse responses
as inputs for counterfactual analysis.

Under policy invariance, the baseline and counterfactual economies share the
equilibrium restrictions \eqref{eq:local-stacked} and differ only in their
shock histories. Let
$\Delta\boldsymbol\varepsilon_t
:=\tilde{\boldsymbol\varepsilon}_t-\boldsymbol\varepsilon_t$
denote the modification to the innovation sequence. Suppose the intervention
begins at date $t$ and is confined to $\mathcal Q=\{0,1,\ldots,q\}$, so that
$\Delta\boldsymbol\varepsilon_s=0$ for $s<t$ and for $s>t+q$. Fix $h\ge 0$.
Since the histories coincide prior to $t$, the deviation at date $t+h$ can be
written as the sum of successive forecast revisions,
\begin{align}
\tilde{\boldsymbol w}_{t+h}-\boldsymbol w_{t+h}
=
\sum_{s=t}^{t+h}
\Bigl(
\mathbb E_{s}[\tilde{\boldsymbol w}_{t+h}-\boldsymbol w_{t+h}]
-
\mathbb E_{s-1}[\tilde{\boldsymbol w}_{t+h}-\boldsymbol w_{t+h}]
\Bigr),
\end{align}
and $\mathbb E_{t-1}[\tilde{\boldsymbol w}_{t+h}-\boldsymbol w_{t+h}]=0$ since
$\Delta\boldsymbol\varepsilon_s=0$ for $s<t$. Each increment reflects the new
information revealed at date $s$. By construction of the filtration and
linearity of equilibrium responses,
\[
\mathbb E_{s}[\tilde{\boldsymbol w}_{t+h}-\boldsymbol w_{t+h}]
-
\mathbb E_{s-1}[\tilde{\boldsymbol w}_{t+h}-\boldsymbol w_{t+h}]
=
\Theta_{t+h-s}\,\Delta\boldsymbol\varepsilon_{s}.
\]
Substitution yields a linear representation of the counterfactual deviation as
the accumulation of period-by-period shock interventions via the same impulse
responses implied by \eqref{eq:local-stacked}. Under the shock-selection
restriction \eqref{cf:shock}, $\Delta\boldsymbol\varepsilon_s$ is nonzero only
in the policy-shock component.

We now specialize to policy interventions. Recall that $z$ is the policy
instrument, a component of $\boldsymbol w$, and that $\delta_\ell$ denotes the intervention in the policy shock at $t+\ell$, see
\eqref{eq:path-restriction}. Recall also that $\psi_{z,h}$ is the impulse
response of $z$ to its own shock at horizon $h$, with $\psi_{z,h}=0$ for $h<0$.
Then the counterfactual deviation of the policy instrument at horizon $h$,
$\tilde z_{t+h}-z_{t+h}$, admits the representation
\begin{equation}
\label{eq:dated-path-dev}
\begin{aligned}
\sum_{\ell=0}^{\min(q,h)}
\Bigl(
\mathbb E_{t+\ell}[\tilde z_{t+h}-z_{t+h}]
-
\mathbb E_{t+\ell-1}[\tilde z_{t+h}-z_{t+h}]
\Bigr)
=
\sum_{\ell=0}^{\min(q,h)}\psi_{z,h-\ell}\,\delta_\ell .
\end{aligned}
\end{equation}
Thus the deviation of the policy instrument at horizon $h$ is the superposition
of the sequential interventions via the corresponding impulse responses.

Stacking the within-window restrictions for horizons $0$ through $q$ yields the
linear system in \eqref{eq:path-restriction}. The same system is obtained from
the SVAR recursion in \eqref{eq:recursive-alg} because both representations are
governed by the same moving-average coefficients for the policy instrument.
Solving the system delivers the sequence of policy-shock interventions
required to implement the targeted within-window path over $\mathcal Q$. Given
this sequence, the counterfactual economy follows by superposition under the
baseline moving-average law of motion. This yields policy-peg impulse responses
and policy-path responses as different evaluations of the same mapping.
Therefore any forward-looking structural model that satisfies the maintained
policy-invariance condition and implies the same moving-average coefficients
induces the identical counterfactual mapping.

\subsubsection{Rule changes and the Lucas critique}
\cite{mckay2023can} study policy rule-change counterfactuals and show that, under
a maintained policy-invariance restriction, a rule perturbation can be
implemented by an appropriately constructed shock sequence. We establish a
complementary equivalence within a fixed linear envelope: a within-window
instrument-path restriction can be implemented by a sequence of policy shocks
that reproduces the targeted instrument path. Combining the two results, our
path counterfactual admits a rule-change rationalization, in the sense that the
same counterfactual instrument path can be represented either as a path
restriction, as its shock implementation, or as a rule change engineered to
deliver that path.

This equivalence is a scope condition rather than a resolution of the Lucas
critique. The critique concerns whether the intervention is perceived as a
perturbation within a stable regime, or as a change that alters private-sector
decision rules and hence the equilibrium law of motion. Results in
\cite{beraja2023semistructural} underscore that reduced-form objects, including
impulse responses, do not by themselves determine how private behavior adjusts
under rule changes. Therefore, absent an additional economic restriction,
transporting impulse responses from the baseline regime to a counterfactual
regime is not warranted. One such restriction is the policy-invariance
condition emphasized by \cite{mckay2023can}, under which policy affects private
behavior only through the expected future path of the instrument. Under this
restriction, the within-window path restriction, its shock implementation, and
its rule-change rationalization are observationally equivalent to private
agents, so the counterfactual mapping is insulated from the Lucas critique
within the maintained linear envelope.

Second, our sequence design can imply interventions that are persistent and
non-marginal within the window. In that case, interpreting the intervention as
a sequence of unexpected shocks may be strained, and the experiment is more
naturally read as a systematic and perceivable policy change. Related concerns are discussed by
\citet{hamilton2004comment}. A standard defense is to restrict attention to
interventions that are small relative to typical disturbances, so that the
sequence remains locally unanticipated and does not trigger a re-optimization of
decision rules, see \citet{kilian2011does}. Practically, this consideration is
most salient for long pegs that span the full adjustment of the economy. When
the peg is temporary and confined to a short window, the scope for endogenous
changes in private sector rules is correspondingly reduced. An alternative is to
adopt a one-shot design, as in \citet{mckay2023can}, which interprets the path
restriction as publicly communicated information at the initial date and hence
implements the restriction through a vector of initial-period shocks rather than a sequence
of shocks. The next section discusses how our LP construction applies to
that design as well.

\section{Local projections for one-shot policy counterfactuals}
\label{sec:oneshot}

This section shows that the local projection framework in Section~\ref{sec3:lp}
also covers one-shot policy counterfactuals in the sense of
\cite{mckay2023can}, in addition to sequential policy-path counterfactuals.

\subsection{One-shot counterfactuals}

The one-shot design modifies the policy shocks only at the event date.
Let $\boldsymbol{\varepsilon}_{z,s}$ denote the vector of policy innovations at
date $s$, let $\boldsymbol{\varepsilon}_{-z,s}$ collect the remaining
structural shocks, and let
$\boldsymbol{\varepsilon}_{s}:=
(\boldsymbol{\varepsilon}_{z,s}',\boldsymbol{\varepsilon}_{-z,s}')'$.
Let $\boldsymbol{\varepsilon}_{-\infty:t-1}$ denote the baseline shock sequence
up to $t-1$. Fix a horizon $h\ge 0$ and define the one-shot counterfactual
continuation by replacing only the date-$t$ policy innovations,
\[
\check{\boldsymbol{\varepsilon}}_{t:t+h}
:=
\bigl(\check{\boldsymbol{\varepsilon}}_{z,t},\boldsymbol{\varepsilon}_{-z,t},
\boldsymbol{\varepsilon}_{t+1},\dots,\boldsymbol{\varepsilon}_{t+h}\bigr),
\]
so that $\boldsymbol{\varepsilon}_{s}$ is unchanged for all $s\ge t+1$ and
$\boldsymbol{\varepsilon}_{-z,t}$ is unchanged at date $t$.
The counterfactual outcome is evaluated under the same structural mapping as in
the baseline economy,
$\check{\boldsymbol w}_{t+h}=\boldsymbol g(\boldsymbol{\varepsilon}_{-\infty:t-1},
\check{\boldsymbol{\varepsilon}}_{t:t+h})$.

The restriction mirrors \eqref{peg}. Over the intervention window, the
restricted instrument path satisfies the same peg constraint on the policy
instruments as in the sequential design,
$\mathbb E\!\left[\check{\boldsymbol z}_{t,\mathcal Q}\mid \varepsilon_{x,t}=1\right]
=
\mathbb E\!\left[\boldsymbol z_{t,\mathcal Q}\mid \varepsilon_{x,t}=0\right]$,
but implementation is now restricted to a single date-$t$ choice of
$\check{\boldsymbol{\varepsilon}}_{z,t}$.

The policy-peg impulse response in the one-shot design is
\begin{align}
\label{irf-peg0}
\beta_{h,\mathcal Q,0}
:=
\mathbb E \!\left[ \check y_{t+h} \mid \varepsilon_{x,t}=1 \right]
-
\mathbb E \!\left[ \check y_{t+h} \mid \varepsilon_{x,t}=0 \right].
\end{align}

Policy path responses are defined by fixing a target path deviation
$\boldsymbol c_{\mathcal Q}$ and comparing the associated one-shot counterfactual
path to its baseline counterpart,
\begin{align}
\label{eq:policy-path-effect0}
\delta_{h,0}(\boldsymbol c_{\mathcal Q})
:=
\check y_{t+h}^{path}(\boldsymbol c_{\mathcal Q})
-
y_{t+h}^{path}.
\end{align}

Exact implementability in a one-shot design reduces to a spanning condition at
date $t$, the contemporaneous policy-innovation vector must span the
instrument-path stack $\boldsymbol z_{t,\mathcal Q}$.

\begin{assumption}[One-shot intervention]
\label{ass:oneshot}
Suppose
\begin{enumerate}[(i)]
\item The intervention operates only through the date-$t$ policy-shock vector,
replacing $\boldsymbol{\varepsilon}_{z,t}$ by $\check{\boldsymbol{\varepsilon}}_{z,t}$,
and leaves all non-policy shocks unchanged.
\item Define $\Sigma_{\mathcal Q,0}:=\mathbb E[\boldsymbol{\varepsilon}_{z,t}\boldsymbol z_{t,\mathcal Q}']$.
Assume that $\Sigma_{\mathcal Q,0}$ is square and nonsingular.
\end{enumerate}
\end{assumption}

Assumption~\ref{ass:oneshot}(\textit{i}), is the one-shot counterpart of
Assumption~\ref{ass:structural-env}(\textit{iii}).
Assumption~\ref{ass:oneshot}(\textit{ii}), imposes exact spanning, the contemporaneous
policy innovations available at date $t$ span $\boldsymbol z_{t,\mathcal Q}$ and
therefore allow the peg to be implemented exactly.
When spanning fails, the Moore--Penrose construction should be read as defining
an approximate one-shot counterfactual, it selects among admissible one-shot
interventions the one that minimizes the least-squares distance to the desired
instrument path.

\subsection{Local projection representation}

Maintain the moving-average envelope and policy invariance in
Assumption~\ref{ass:structural-env}, parts i and ii.
Identification in the one-shot design uses the shock of interest and the
contemporaneous policy-innovation vector at date $t$, which instruments the
post-treatment stack $\boldsymbol z_{t,\mathcal Q}$.
For each horizon $h\ge 0$,
\begin{align}
\label{eq:lp-oneshot}
y_{t+h}
=
\beta_{h,\mathcal Q,0}\,\varepsilon_{x,t}
+
\boldsymbol\gamma_{h,\mathcal Q,0}'\boldsymbol z_{t,\mathcal Q}
+
u_{t,h,\mathcal Q,0},
\end{align}
with $\mathbb E[\mathcal E_{t,0} u_{t,h,\mathcal Q,0}]=0$ and
$\mathcal E_{t,0}:=(\varepsilon_{x,t},\boldsymbol{\varepsilon}_{z,t}')'$.
Recall the LP regressor stack
$X_{t,\mathcal Q}:=(\varepsilon_{x,t},\boldsymbol z_{t,\mathcal Q}')'$.
Define
$\Bar{\Sigma}_{\mathcal Q,0}:=\mathbb E[\mathcal E_{t,0}X_{t,\mathcal Q}']$
and
$\Bar{\Gamma}_{h,0}:=\mathbb E[\mathcal E_{t,0}y_{t+h}]$.
With $\lambda$ selecting the coefficient on $\varepsilon_{x,t}$ in
$X_{t,\mathcal Q}$, the one-shot policy-peg coefficient is identified as
\begin{align}
\label{eq:beta-gamma-oneshot}
\beta_{h,\mathcal Q,0}^{LP}
=
\lambda'(\Bar\Sigma_{\mathcal Q,0})^{-1}\Bar\Gamma_{h,0}.
\end{align}

For policy-path responses, given a path deviation vector $\boldsymbol c_{\mathcal Q}$,
the induced outcome deviation admits the linear representation
\begin{align}
\delta_{h,0}^{LP}(\boldsymbol c_{\mathcal Q})
=
(\boldsymbol\gamma_{h,\mathcal Q,0}^{LP})'\boldsymbol c_{\mathcal Q},
\end{align}
where
$\boldsymbol\gamma_{h,\mathcal Q,0}^{LP}
=
\Sigma_{\mathcal Q,0}^{-1}\Gamma_{h,0}$,
with
$\Gamma_{h,0}:=\mathbb E[\boldsymbol{\varepsilon}_{z,t}y_{t+h}]$
and $\Sigma_{\mathcal Q,0}$ as in Assumption~\ref{ass:oneshot}.

The practical obstacle for one-shot counterfactuals is dimensionality.
Unless researchers can isolate a sufficiently rich vector of contemporaneous
policy and news shocks at date $t$ to span the targeted window $\mathcal Q$,
exact implementation over $\mathcal Q$ is infeasible.
In much of the proxy-shock literature on monetary policy, the available
contemporaneous shock vector is low-dimensional, often limited to a monetary
policy shock and a central bank information shock, see \cite{inoue2021new} and
\cite{jarocinski2024estimating}, among others.

\subsection{LP equivalence in one-shot counterfactual}

Under the linear moving-average envelope in Section~\ref{subsec:svma-envelope},
the one-shot counterfactual admits a closed-form representation that parallels
the sequential design.
For the intervention window $\mathcal Q$, let
$\Delta_{0}:=\check{\boldsymbol{\varepsilon}}_{z,t}-{\boldsymbol{\varepsilon}}_{z,t}$
denote the date-$t$ policy intervention in Assumption~\ref{ass:oneshot}.
\begin{align}
\Psi_0(\mathcal Q)\,\Delta_{0}+\boldsymbol z_{t,\mathcal Q}
=
\check{\boldsymbol z}_{t,\mathcal Q},
\end{align}
where $\Psi_0(\mathcal Q):=\partial \boldsymbol z_{t,\mathcal Q}/\partial \boldsymbol{\varepsilon}_{z,t}'$.
In contrast to the sequential design, $\Psi_0(\mathcal Q)$ need not be lower
triangular, since the entire window is affected through the contemporaneous
policy-shock vector at date $t$.

In the policy-peg counterfactual, the intervention $\Delta_0^{peg}$ is chosen to
offset the unrestricted instrument-path response, so it satisfies
$\Psi_0(\mathcal Q)\Delta_{0}^{peg}=-\boldsymbol{\theta}_z(\mathcal{Q})$,
which parallels \eqref{eq:mean-shift-mq} in the sequential design.
Under Assumption~\ref{ass:oneshot}, part ii, this equation has a unique
solution. The one-shot policy-peg impulse response then equals the sum of the
direct effect of the shock of interest and the contemporaneous effect of the
policy intervention,
\begin{align}
\beta_{h,\mathcal Q,0}
=
\theta_{y,h} + \boldsymbol{\psi}_{y,h,0}' \Delta_{0}^{peg},
\end{align}
where $\boldsymbol{\psi}_{y,h,0}$ denotes the response of the outcome to the
date-$t$ policy-shock vector.

Similarly, for policy-path counterfactuals, $\Delta_0^{path}$ is chosen to hit
the target deviation $\boldsymbol c_{\mathcal Q}$, so it satisfies
$\Psi_0(\mathcal{Q})\Delta_0^{path}=\boldsymbol c_{\mathcal{Q}}$,
which parallels \eqref{eq:m-diff}. The corresponding path response is
\begin{align}
\delta_{h,0}(\boldsymbol c_{\mathcal{Q}})
=
\boldsymbol{\psi}_{y,h,0}' \Delta_0^{path}.
\end{align}

The preceding identities show that the LP coefficients induce the same one-shot policy-peg and policy-path responses as those implied by the linear moving-average envelope.

\begin{theorem}[Local projection identification of the one-shot policy counterfactuals]
\label{theo:oneshot}
Suppose Assumption~\ref{ass:structural-env}(\textit{i})-(\textit{ii}) and
Assumption~\ref{ass:oneshot} hold. Then
$\beta_{h,\mathcal Q,0}=\beta_{h,\mathcal Q,0}^{LP}$
and
$\delta_{h,0}(\boldsymbol c_{\mathcal Q}) = \delta_{h,0}^{LP}(\boldsymbol c_{\mathcal Q})$.
\end{theorem}

The proof is given in Appendix \ref{proof:oneshot}.

\section{Scope and interpretation}
\label{sec:miscel}

\subsection{Costs and scope of local projections}

Local projections enter our analysis primarily through identification. Under
policy invariance and a linear SVMA envelope, a finite-horizon policy-path
restriction is pinned down by a small set of impulse-response objects together
with the imposed path deviation. This reduction yields a transparent
interpretation. The counterfactual is a regression object indexed by policy
relevant dynamic responses, which clarifies which features are identified and
which restrictions are maintained.

The same structure streamlines implementation. Rather than re-estimating a
large collection of horizon-specific and window-specific coefficients, one can
estimate the relevant impulse responses and then evaluate the counterfactual by
plug-in. This separation is useful in macroeconomic samples, where parameter
proliferation is a first-order concern. As the horizon $h$ and the intervention
length $q$ grow, a direct LP implementation accumulates sampling error across
both dimensions and precision deteriorates. Working through impulse responses
and an explicit mapping curbs this accumulation and delivers numerically stable
counterfactual paths in limited samples.

The scope restriction tightens in nonlinear and state-dependent environments.
When propagation varies with an underlying state, impulse responses are
state-specific, and a policy-path restriction implemented through repeated
interventions can shift the distribution of future states. The sequential
design is particularly exposed because maintaining a peg typically requires
sustained offsets that may induce repeated, endogenous state transitions. In
such settings, impulse responses from a linear envelope need not transport to
the counterfactual path, and credible policy evaluation typically requires a
model that makes the state dependence and transition mechanism explicit. The LP
implementation developed here should therefore be read as a tractable device
for counterfactual analysis in linear environments, not as a substitute for
nonlinear policy evaluation.

\subsection{Plausibility of policy counterfactuals}

The representation also provides a practical diagnostic for plausibility. It
separates two margins. The first margin concerns the transmission objects used
to construct the counterfactual. The second margin concerns the admissibility
of the intervention sequence required to implement the restriction.

The first margin is inherited from the impulse-response inputs. Counterfactual
paths are obtained by mechanically combining estimated responses with a
policy-path restriction, so they inherit the identification, specification, and
sampling assumptions behind those responses. In applications, impulse-response
estimates can vary across designs, across samples, and across identification
schemes. Under our construction, such variation maps directly into variation in
the implied counterfactual paths.

The second margin concerns implementation under policy invariance. We view the
counterfactual as a temporary modification of the policy-shock sequence that
enforces the instrument restriction while keeping transmission fixed. The
practical question is whether the implied intervention sequence is admissible
in magnitude and persistence. A key advantage of our mapping is transparency,
it delivers the implied shocks explicitly, so applications can report their
size and persistence alongside the counterfactual path.

One-shot counterfactuals face a different bottleneck. They use a finite vector
of initial period shocks to approximate a multi-horizon policy-path
restriction. Unless the design identifies a sufficiently rich set of
contemporaneous policy and news shocks, exact matching of a long-horizon peg is
generically infeasible, so the exercise is an approximation problem. The
weighting across horizons is then best viewed as a researcher-chosen loss
function that ranks horizons, rather than as an object pinned down by the data.
The one-shot design is also compelling only when agents plausibly observe and
internalize the future scheduled interventions at the initial date. When this informational
premise is doubtful, the sequential counterfactual provides a sharper benchmark.

\section{Empirical studies of monetary policy counterfactuals}
\label{sec6:counterfactual monetary policy}

This section implements two sequential policy counterfactuals in the linear,
policy-invariant environment maintained above. The first shuts down the endogenous policy-rate response along the propagation
of an oil-supply news shock by fixing the federal funds rate at its baseline
path over a prespecified window. The second constructs a retrospective
policy-path counterfactual for the post-pandemic tightening episode by imposing
an earlier liftoff of the policy rate.

\subsection{Policy-peg impulse responses to oil price shocks}
\label{subsec:sz}

This subsection studies sequential policy-rate pegs following an oil-supply news
shock. The counterfactual holds the federal funds rate at its baseline path over
a window and lets all non-policy margins adjust endogenously. Comparing the
unrestricted oil-shock response to the peg response isolates the contribution
of the systematic monetary policy reaction in the propagation of the oil
disturbance.

The construction uses two identified shocks with distinct roles. The oil-supply
news shock is identified using the oil-supply news surprise series of
\cite{kanzig2021macroeconomic}.\footnote{The oil-supply news surprise data are
obtained from Diego Känzig's website.} The peg is implemented as a sequence of monetary policy shocks that offsets the endogenous responses of the
federal funds rate induced by the oil shock. These monetary policy shocks are
identified using high-frequency monetary policy surprises from
\cite{bauer2023reassessment}. The sample comprises monthly data on log WTI oil
price, log industrial production, CPI inflation, the federal funds rate, and the
excess bond premium from February 1988 to December 2019.\footnote{The excess
bond premium was introduced by \cite{gilchrist2012credit}. We use the updated
series provided in the FEDS Notes by \cite{favara2016updating_feds}.}

Local projections serve here to state the peg estimand and fix its
interpretation, rather than as the primary estimation device. For a peg length
$q$ and horizon $h$, the policy-peg regression can be
written as
\begin{align}
\label{eq:lp-peg-empirical}
y_{t+h}
=
\beta_{h,\mathcal Q}\,oilshock_t
+
\boldsymbol\gamma_{h,\mathcal Q}'\, {ffr}_{t,\mathcal Q}
+
u_{t,h,\mathcal Q}.
\end{align}
Equation \eqref{eq:lp-peg-empirical} makes clear how the counterfactual
restriction enters through the within-window policy path, and it clarifies the
object targeted by a horizon-by-horizon peg regression. For estimation, we
implement the peg through the explicit mapping in \eqref{cir:post}. The
practical reason is sample usage. A direct LP implementation conditions on the
policy path $(ffr_t,\ldots,ffr_{t+q})$ at each horizon, so the effective sample
shrinks quickly as $(h,q)$ increase, which is especially costly for long pegs.
We therefore proceed in two steps. First, we estimate the impulse responses to
the oil-supply news shock and to the monetary policy shock using a structural
VAR with the five variables and twelve lags, augmented with the corresponding
external instruments. Second, we plug these estimated impulse responses into
\eqref{cir:post} to obtain the policy-peg impulse response. This approach
retains the counterfactual interpretation highlighted by
\eqref{eq:lp-peg-empirical} while avoiding the loss of observations implied by
horizon-by-horizon peg regressions. When sample size permits and the intervention window is short, the same object can be estimated directly by stacked LP--IV using the proxy stack.

\begin{figure}[ht!]
    \centering
    \includegraphics[width=0.9\textwidth]{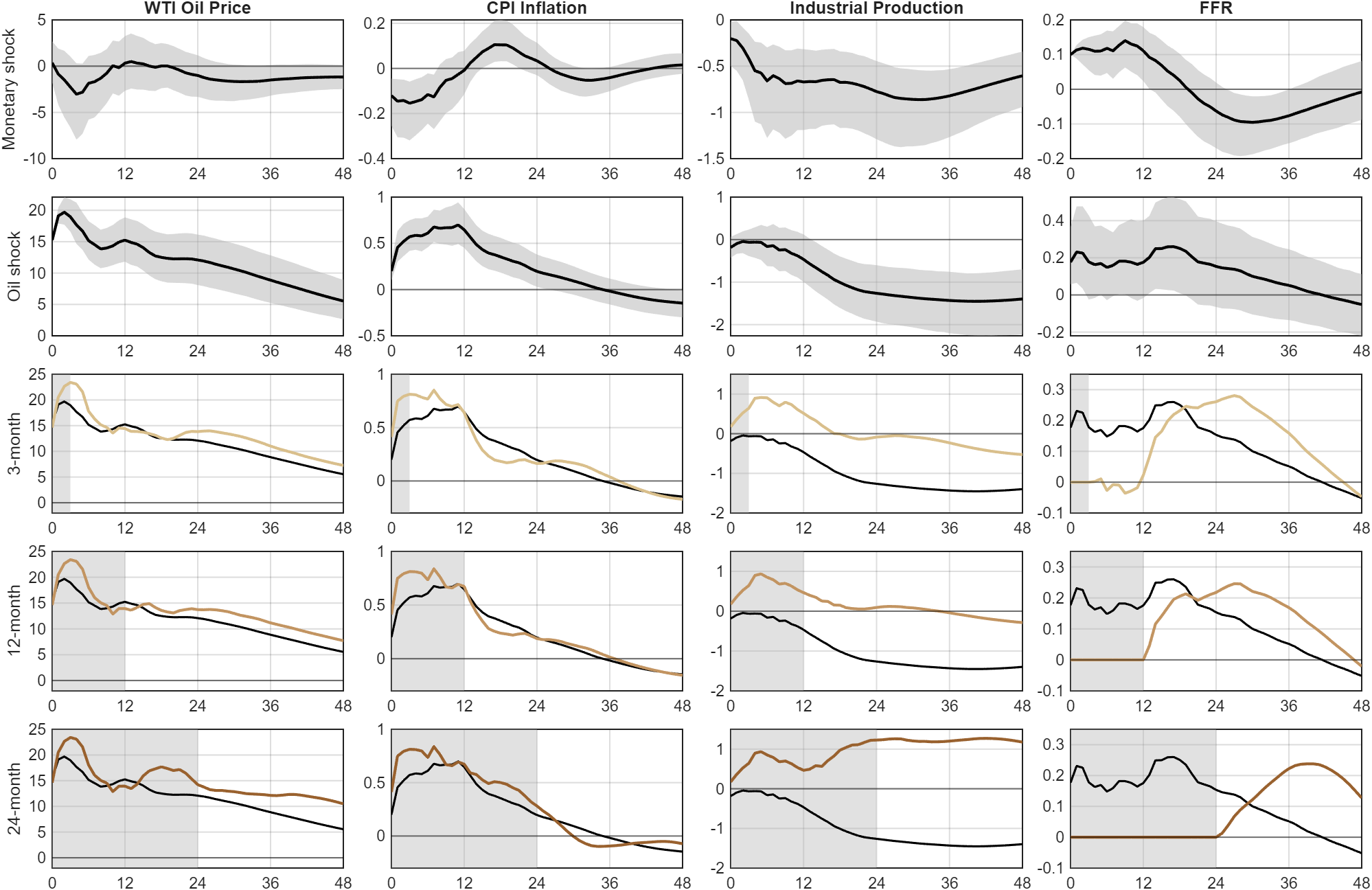}
    \begin{minipage}{0.9\textwidth}
    \caption{Policy-peg impulse responses to an oil-supply news shock under a sequential design. \footnotesize
    The figure reports impulse responses over monthly horizons
    $h=0,1,\ldots,48$ for, from left to right, WTI oil price, CPI inflation,
    industrial production, and the federal funds rate, denoted FFR. The first
    row reports responses to a 10 bps contractionary monetary policy shock. The
    second row reports responses to a one-standard-deviation oil-supply news
    shock, normalized so that the impact response of WTI is about 15\%. Shaded
    bands in the first two rows are 68\% confidence intervals computed from
    5{,}000 block bootstrap replications with block size 24. Rows 3 to 5 report
    policy-peg counterfactual responses to the oil-supply news shock under peg
    windows $q=3$, $q=12$, and $q=24$ months. In these rows, the shaded region
    marks the intervention window during which the FFR is fixed at its baseline
    path. The black line reproduces the unrestricted oil-shock impulse response and the
    colored line reports the corresponding peg counterfactual. The required policy intervention for this peg, rounded to 5~bps increments, is
$(-20,-5,0,5)$ over months $0$--$3$, $(0,0,5,0,0,5,0,0,-5)$ over months $4$--$12$,
and $(-5,-5,0,0,-5,0,0,0,0,0,0,0)$ over months $13$--$24$.
}
    \label{fig:sz}
    \end{minipage}
\end{figure}

Figure~\ref{fig:sz} reports the unrestricted propagation of the monetary policy
shock and the oil-supply news shock, together with the implied policy-peg
counterfactuals.\footnote{Policy-peg impulses are nonlinear functionals of estimated impulse responses. As the intervention window and projection horizon increase, sampling uncertainty accumulates, so confidence bands are typically wide and less informative for inference. Bands can nevertheless be constructed by propagating bootstrap draws of the impulse responses through the counterfactual mapping. Our focus here is on the point estimate relative to the realized path.} The first row shows that a 10 bps contractionary monetary
policy shock lowers WTI, with a trough of roughly 3--4\% within the first few
months. CPI inflation falls on impact by about 0.1--0.2\% and then reverts
gradually. Industrial production contracts persistently, with responses around
$-0.6$ to $-1.0$\% over horizons between one and three years.

The second row shows that a one-standard-deviation oil-supply news shock raises
CPI inflation quickly, reaching about 0.6--0.8\% in the first year before
declining gradually thereafter. Real activity falls with delay and persistence.
Industrial production moves below baseline within the first year and reaches
around $-1.2$ to $-1.5$\% at horizons between two and four years. The federal
funds rate tightens endogenously, increasing by about 20 bps during the first
two years, consistent with an active stabilization response to oil-driven
inflationary pressure.

Rows three to five report peg counterfactuals. Shutting down the endogenous policy
response changes both the timing and the magnitude of stabilization, and the
effects increase with the peg length. Under all three pegs, CPI inflation is
higher during the intervention window than under the unrestricted response.
Over the first year, the peg raises inflation by roughly 0.15--0.30\%. The WTI response is also higher in the near term, by about 5\% on impact, suggesting that part of the observed oil--inflation comovement reflects the counterfactual policy-peg environment in which the policy rate is held fixed rather than cut as in the data. Industrial production is likewise higher by about 0.3--1\% over the first year, indicating a nontrivial real activity expansion under the peg.

When the peg expires, the policy rate adjusts with delay and displays catch-up
tightening. By construction, the federal funds rate is flat during the window,
and the subsequent tightening is shifted into later horizons as the peg
lengthens. In the 12-month and 24-month designs, the post-peg increase is on
the order of 20--30 basis points. The counterfactual therefore re-times, rather
than removes, the monetary response to the oil shock, and this timing shift is
central for the ensuing inflation and output paths.

The most pronounced differences arise for real activity. Under the unrestricted
response, industrial production is about $-1.2$\% at the two-year horizon. Under
a 12-month peg, the corresponding path is close to zero at that horizon, an
attenuation of roughly 1\%. Under a 24-month peg, industrial production turns
positive and remains elevated. By three years, the peg path is about 1\% above
baseline, while the unrestricted path is around $-1.4$\%, a gap of roughly
2.5\%. The widening gap as the peg window extends indicates a sizable role for
systematic monetary tightening in the medium-run output contraction following
oil shocks.

The inflation trade-off varies with the peg duration. Under a 3-month peg,
inflation is higher over the first year and then converges back toward the
unrestricted path, with a modest dip at intermediate horizons. This dip follows
from a sign reversal in the inflation responses reported in the first row of
Figure~\ref{fig:sz}: CPI inflation falls on impact, but the estimated response
to a contractionary monetary policy shock turns positive at intermediate
horizons, so the offsetting shocks used to hold the federal funds rate fixed
pull peg inflation below the unrestricted path over that range. The 12-month peg
delivers a similar profile because the 3-month peg already keeps the implied
funds-rate deviations close to zero through horizon twelve, so extending the peg
to twelve months requires only modest additional intervention. By contrast, the
24-month peg sustains higher inflation throughout the intervention window, with
convergence toward the unrestricted response only after the peg ends. Taken together, the three policy-peg windows remove endogenous monetary policy
responses after the oil price shock, period by period, and show how their
cumulative effect governs the resulting inflation--output trade-off.

\subsection{Policy-path effects in the post-pandemic era}
\label{subsec:postpandemic_path}

This subsection constructs a retrospective counterfactual for the post pandemic
tightening episode. We take the realized sequence as the baseline and impose a
policy-rate path with an earlier liftoff over a prespecified window. The
counterfactual is defined by the sequence of monetary policy shocks that closes
the gap between the realized and imposed federal funds rate within the window,
given the estimated impulse responses of the policy rate to the identified
policy shock. The object of interest is the magnitude, timing, and persistence of
the implied deviations from realized outcomes.

The evaluation window is January~2021--April~2025. The monthly dataset includes
the effective federal funds rate, the unemployment rate, CPI inflation, real
GDP, and the excess bond premium. The federal funds rate, unemployment, and CPI
inflation are taken from FRED. Monthly real GDP is measured by S\&P Global
Market Intelligence's monthly GDP index. Throughout, reported gaps are expressed in
percentage points for the federal funds rate, unemployment, and inflation, and
in percent for real GDP, relative to the realized path. Oil prices are omitted
by design. This choice aligns with monetary policy identification exercises
that trace policy-shock effects without conditioning on oil-price dynamics, see
\cite{gertler2015monetary}, \cite{nakamura2018high}, and
\cite{bauer2023reassessment}.

\begin{figure}[ht]
    \centering
    \includegraphics[width=0.9\textwidth]{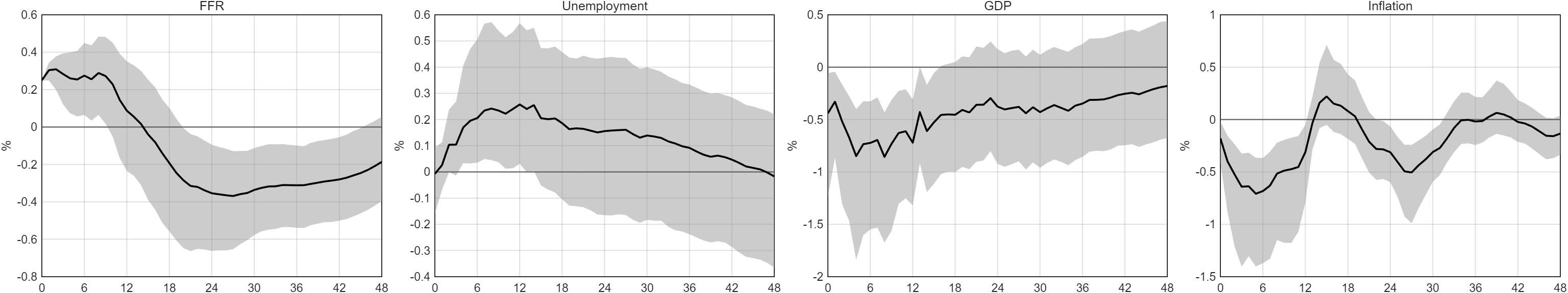}
    \begin{minipage}{0.9\textwidth}
    \caption{Impulse responses to a 25-basis-point monetary policy shock.
    \footnotesize
    The figure reports impulse responses from a structural VAR identified
    with a high-frequency external instrument, the orthogonal monetary policy
    shock of \cite{bauer2023reassessment}. The VAR includes the FFR, the
    unemployment rate, real GDP, CPI inflation, and the excess bond premium.
    The lag order is fifteen. The model is estimated over 1992:01--2020:02,
    338 monthly observations. Shaded bands are 68\% confidence intervals
    obtained from a block bootstrap with 5{,}000 replications and block length
    24 months.}
    \label{fig:irf}
    \end{minipage}
\end{figure}

Local projections provide a useful representation of the policy-path estimand.
For an intervention window of length $q$ and horizon $h$, a policy-path
regression can be written as
\begin{align}
\label{eq:lp-path-empirical}
    y_{t+h}
    =
    \gamma_{h,\mathcal{Q}}'\,\bigl(ffr_t,\ldots,ffr_{t+q}\bigr)
    +
    v_{t,h,\mathcal{Q}}.
\end{align}
Equation \eqref{eq:lp-path-empirical} makes explicit that the estimand is a
mapping from the within-window policy path to outcomes. In implementation, we
use the explicit mapping in \eqref{cir:post}. We first estimate impulse
responses to an identified monetary policy shock, then apply \eqref{cir:post}
to translate an imposed deviation in the federal funds rate into counterfactual
trajectories for unemployment, real GDP, and inflation. The impulse responses are obtained from a structural VAR identified with a
high-frequency external instrument, the orthogonal monetary policy shock of
\cite{bauer2023reassessment}. The VAR includes the federal funds rate, the
unemployment rate, real GDP, CPI inflation, and the excess bond premium. The
model is estimated over 1992:01--2020:02 with a matched subsample for the
instrument, and Figure~\ref{fig:irf} reports the resulting responses.

We define the policy-path experiments around the onset of post-pandemic tightening. The first increase in the target range after the pandemic occurred on March 2022. In the effective-rate series, however, the federal funds rate begins to move away from the zero region in February~2022, reaching about 0.2\%. We take February~2022 as the onset of liftoff in the effective rate and consider counterfactuals in which liftoff starts three months earlier, in November~2021. We vary the intervention length. A scenario labeled as a $q$-month intervention refers to a window whose terminal month is $q$ months after the beginning month, with both endpoints included. In the notation above, this corresponds to $\mathcal Q=\{0,1,\dots,q\}$ and a $q\!+\!1$ element vector of policy shocks.

In the three-month scenario, the window runs from November~2021 through February~2022, so the intervention is implemented by four shocks, November~2021 through February~2022, and we shift the realized policy-rate path over February--May~2022 back to November~2021--February~2022. In the six-month scenario, the window runs from November~2021 through May~2022, so the intervention is implemented by seven shocks, November~2021 through May~2022, and we shift the realized policy-rate path over February--August~2022 back to November~2021--May~2022. After the window ends, the counterfactual policy rate evolves according to the estimated dynamics implied by the imposed shock sequence. Counterfactual paths for unemployment, real GDP, and inflation are computed from the common start date, November~2021.

\begin{figure}[ht!]
    \centering
    \includegraphics[width=0.9\textwidth]{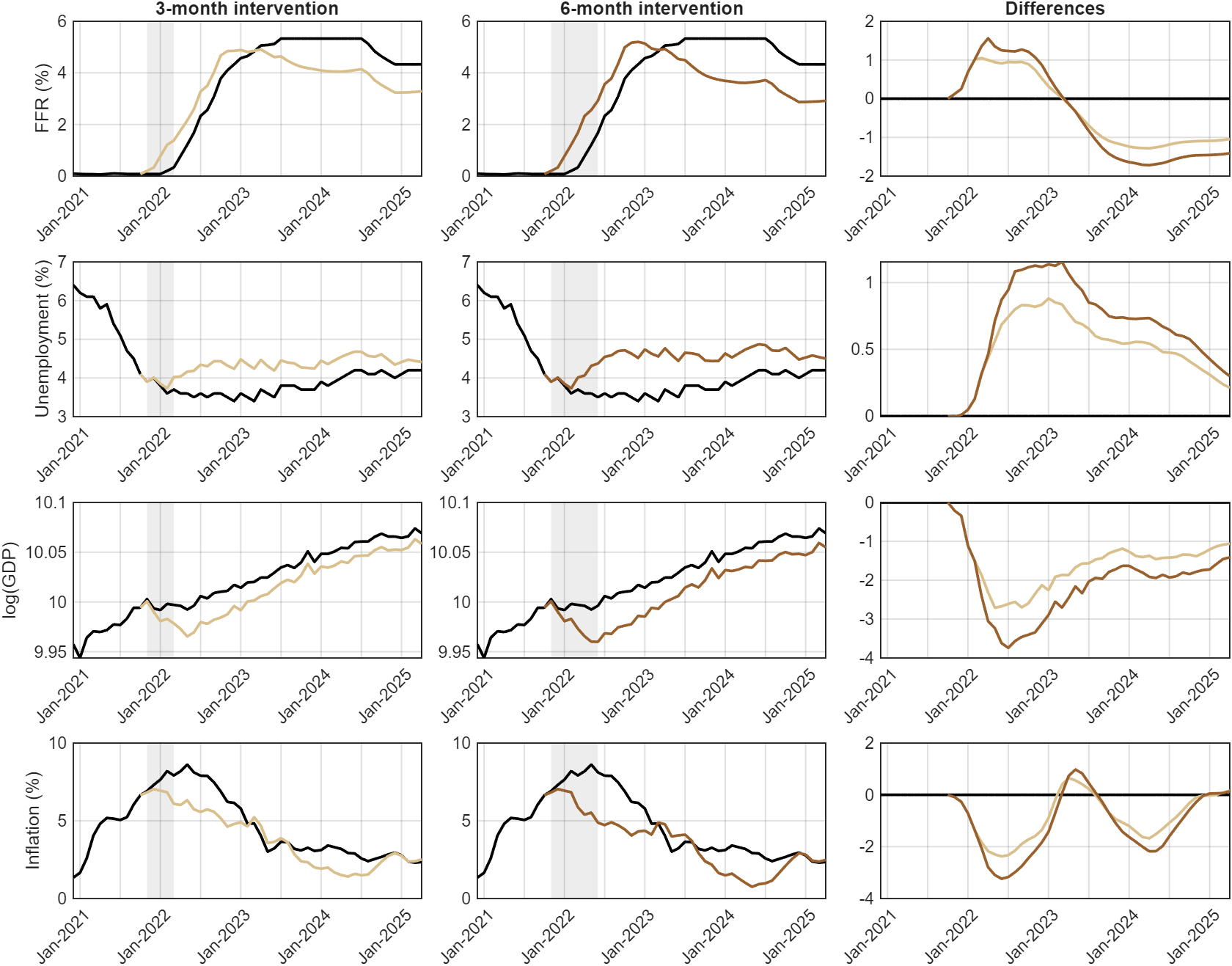}
    \begin{minipage}{0.9\textwidth}
    \caption{Hypothetical policy paths and implied outcome gaps in the post-pandemic era.
    \footnotesize
    The black line reports the historical series. The counterfactual policy path
under the three-month intervention is shown in light ochre, and the
counterfactual under the six-month intervention is shown in dark amber. The
shaded band marks the intervention window beginning in November~2021. The left
panel plots the three-month intervention (November~2021 to February 2022), the middle panel plots the six-month
intervention (November~2021 to May 2022), and the right panel plots the corresponding deviations from the
historical series implied by the two policy-path responses. For the three-month scenario, the required policy shock interventions are approximately
$(10,10,40)$~bps, rounded to the nearest $5$~bps. For the six-month scenario,
the required shocks are approximately $(10,10,40,25,30,20)$~bps.
}
    \label{fig:path}
    \end{minipage}
\end{figure}

Figure~\ref{fig:path} summarizes the imposed policy-rate paths and the implied
outcome gaps relative to history. Under the three-month scenario, unemployment
rises by about 0.5\% at its peak in early 2023 and then
converges toward the realized path by 2025. Real GDP falls by up to about 2\%
around mid-2022 and remains roughly 1\% below the realized path in 2025. CPI
inflation declines by about 2\% in mid-2022 and about 1\% in early 2024, and it returns close to the realized path in
early 2025.

Under the six-month scenario, the imposed deviation in the policy path is more
persistent and the implied comovement between disinflation and real activity is
stronger. Unemployment increases by up to about 1.3\% before
gradually converging back. Real GDP falls by up to about 4\%, and the shortfall
remains sizable, still around 2\% below the realized path by 2025. CPI inflation
declines more sharply, by up to about 3.5\% in mid-2022 and
about 2.5\% around mid-2024, before approaching the realized
path in 2025. Comparing the two windows shows that a longer earlier liftoff
produces larger and more persistent disinflation under the estimated 
impulse-response mapping, alongside a deeper and more sustained contraction in
real activity.

Our method also recovers the policy-shock sequence that implements each imposed path. Using the estimated impulse response of the federal funds rate as the shock-to-path operator, we back out the monthly shocks that close the gap between the historical and hypothetical federal funds rate over the intervention window. Because the three-month and six-month interventions share the same first three months, the inferred shocks coincide over that common segment. While unusually large shocks could strain policy invariance in the sense of \cite{lucas1976econometric} and raise concerns about the maintained invariance restriction, our counterfactual does not posit unprecedented magnitudes. It reuses a policy-intervention episode observed in the data and shifts its timing three months earlier to enforce the imposed path, in the similar spirit of historical-decomposition based scenarios as in \cite{baumeister2014oil}, see also the discussion in Section~4 of \cite{kilian2017structural}.

This exercise is retrospective and locally linear. It takes the estimated impulse-response mapping as fixed and asks what sequence of monetary policy interventions would have been required to deliver an earlier liftoff, conditional on the realized sequence of shocks. The policy implication is quantitative and trade-off based: under the estimated dynamics, advancing liftoff would have reduced inflation earlier and by a nontrivial margin, at the cost of a measurable increase in unemployment and a persistent output shortfall, with magnitudes rising with the length of the intervention window. Given a researcher-specified welfare function, the resulting counterfactual paths can be ranked and used to characterize the policy that minimizes the loss within the class of feasible short-rate paths considered here.

\subsection{Empirical implications}

The two counterfactual paths in this section are functions of a finite collection of impulse-response estimates. Conditional on these inputs, the mapping is explicit and delivers a unique counterfactual path. It follows that disagreement about counterfactual magnitudes is, in substance, disagreement about the impulse-response estimates that enter the mapping.

This structure yields a simple diagnostic for interpretation. The counterfactual can be written as a horizon-dependent functions of impulse responses, so one can identify the horizons that account for the bulk of the counterfactual movement and then inspect the corresponding impulse-response estimates. The empirical question is whether the decisive responses, in magnitude, persistence, and timing, are credible under the maintained identifying restrictions and the economic restrictions invoked for interpretation, given sampling uncertainty in the estimated impulse responses.

Because impulse responses are identification outcomes, they can vary across credible designs. They depend on the identification method, the sample period, and the model specification used for estimation, and these margins can materially affect estimated response profiles, see \cite{ramey2016macroeconomic}. Accordingly, applied work should treat the impulse-response inputs as a sensitivity margin, recompute the same counterfactual under a small set of alternative impulse-response estimates, for example varying lag specification and subsamples to probe stability, and report the implied envelope of counterfactual paths. When the envelope is tight, the interpretation is more persuasive. When it is not, the dispersion pinpoints the horizons and impulse-response components that are decisive for the exercise.

\section{Conclusion}
\label{sec9:concl}

This paper develops a local projection framework for macroeconomic counterfactuals
that constrain a policy instrument over a finite horizon. We study two policy
objects, policy-peg impulse responses and policy-path effects. Both objects
restrict the instrument path on a prespecified intervention window, while
allowing all non-policy margins to adjust endogenously.

Our main result establishes equivalence between the LP estimands and the
structural counterfactual objects under a linear moving-average envelope and a
maintained policy invariance restriction. In population, the counterfactual
outcomes are identified as local projection coefficients. Moreover, any
empirical SVAR implementation or forward-looking structural model that implies
the same moving-average representation for observables delivers the same
counterfactual outcomes. In particular, the local projection estimand induces an
explicit counterfactual mapping. A finite collection of impulse responses
suffices to recover the sequence of policy interventions that implements
the imposed restriction and the implied counterfactual outcomes.

We maintain a policy invariance restriction. Two conditions
can be thought of as supporting this restriction. First, one may impose structural conditions on the underlying model so that the
private-sector block is held fixed with respect to the policy intervention and
the intervention operates only through the policy instrument path. Second, the policy shock used to
implement the restriction retains its interpretation as an unexpected policy movement. This condition is more demanding for sequential designs, because repeated
large, same-sign interventions may shift expectations, so the exercise is most
credible when the intervention is temporary and moderate. The one-shot design
mitigates this concern by construction, but it treats private agents as
observing the full future restriction at the event date.

Two monetary applications illustrate the empirical content of the framework. In
the oil-supply shock exercise, imposing a short-window policy-rate peg, which
shuts down the policy transmission channel within the window, raises
inflation on impact and yields a milder medium-run contraction in activity. In
the post-pandemic liftoff exercise, alternative policy-rate paths imply
progressively larger disinflation and a sharper contraction in real activity as
the intervention window lengthens. Constructing the policy-shock sequence
that implements each imposed path restriction makes the intervention explicit and provides
a feasibility check under the historical equilibrium.

\appendix

\section{Appendix: proofs}
\subsection{Theorem \ref{expostprop}}
\label{proof:expost}

\begin{proof}
Recall $\beta_{h,\mathcal Q}^{LP}=\lambda'\bar\Sigma_{\mathcal Q}^{-1}\bar\Gamma_{h,\mathcal Q}$.
Partition $\bar\Sigma_{\mathcal Q}$ conformably with $\bigl(\varepsilon_{x,t},\,\boldsymbol z_{t,\mathcal Q}'\bigr)'$.
Under the orthogonality of $\varepsilon_{x,t}$ and $\varepsilon_{z,t+\ell}$ for all $\ell$,
block inversion gives $\lambda'\bar\Sigma_{\mathcal Q}^{-1}
=\mathbb E[\varepsilon_{x,t}^2]^{-1}\bigl[1,\,-\mathbb E[\varepsilon_{x,t}\boldsymbol z_{t,\mathcal Q}']\Sigma_{\mathcal Q}^{-1}\bigr]$ .
Since $\bar\Gamma_{h,\mathcal Q}=\bigl(\mathbb E[\varepsilon_{x,t}y_{t+h}],\,\Gamma_{h,\mathcal Q}'\bigr)'$, substitution yields
\begin{align}
\beta_{h,\mathcal Q}^{LP}
=\theta_{y,h}-\boldsymbol{\theta}_z(\mathcal Q)'\Sigma_{\mathcal Q}^{-1}\Gamma_{h,\mathcal Q},
\end{align}
using $\theta_{y,h}=\mathbb E[\varepsilon_{x,t}^2]^{-1}\mathbb E[\varepsilon_{x,t}y_{t+h}]$ and
$\boldsymbol\theta_z(\mathcal Q)=\mathbb E[\varepsilon_{x,t}^2]^{-1}\mathbb E[\varepsilon_{x,t}\boldsymbol z_{t,\mathcal Q}]$ by the impulse response definition and structural shocks are orthogonal.

To conclude $\beta_{h,\mathcal Q}^{LP}=\beta_{h,\mathcal Q}$, it remains to rewrite $\Sigma_{\mathcal Q}^{-1}\Gamma_{h,\mathcal Q}$ in the form used in the definition of $\beta_{h,\mathcal Q}$. Let $\Sigma_z := I_{q+1}\otimes \mathbb E[\varepsilon_{z,t}^2]$ and use the impulse-response definition
$\Psi(\mathcal Q)' = \Sigma_z^{-1}\Sigma_{\mathcal Q}$ and
$\boldsymbol\psi_{y,h,\mathcal Q}=\Sigma_z^{-1}\Gamma_{h,\mathcal Q}$.
Then
\begin{align}
\label{4.9}
\Sigma_{\mathcal{Q}}^{-1}\Gamma_{h,\mathcal{Q}}
=(\Sigma_z^{-1}\Sigma_{\mathcal{Q}})^{-1} \Sigma_z^{-1}\Gamma_{h,\mathcal{Q}}
=\Psi(\mathcal{Q})^{-1'}\boldsymbol\psi_{y,h,\mathcal Q}.
\end{align}
Substituting this identity into the previous equation gives $\beta_{h,\mathcal Q}^{LP}=\beta_{h,\mathcal Q}$.

Finally, $\boldsymbol\gamma_{h,\mathcal Q}^{LP}=\Sigma_{\mathcal Q}^{-1}\Gamma_{h,\mathcal Q}$ by LP equation \eqref{eq:lp-path-reg}. Define the structural path coefficient vector
\begin{align}
\boldsymbol\gamma_{h,\mathcal Q}
:=
\Psi(\mathcal Q)^{-1'}\,\boldsymbol\psi_{y,h,\mathcal Q},
\end{align}
so that $\delta_h(\boldsymbol c_{\mathcal Q})=\boldsymbol\psi_{y,h,\mathcal Q}'\Psi(\mathcal Q)^{-1}\boldsymbol c_{\mathcal Q}=\boldsymbol\gamma_{h,\mathcal Q}'\boldsymbol c_{\mathcal Q}$ follows \eqref{eq:delta-realization}. The preceding algebra implies $\Sigma_{\mathcal Q}^{-1}\Gamma_{h,\mathcal Q}=\Psi(\mathcal{Q})^{-1'}\boldsymbol\psi_{y,h,\mathcal Q}$, hence $\boldsymbol\gamma_{h,\mathcal Q}^{LP}=\boldsymbol\gamma_{h,\mathcal Q}$. Therefore,
\begin{align}
\delta_h^{LP}(\boldsymbol c_{\mathcal Q})
&=
(\boldsymbol\gamma_{h,\mathcal Q}^{LP})'\boldsymbol c_{\mathcal Q}
=
\boldsymbol\psi_{y,h,\mathcal Q}'\Psi(\mathcal Q)^{-1}\boldsymbol c_{\mathcal Q}
=
\delta_h(\boldsymbol c_{\mathcal Q}).
\end{align}
This completes the proof.
\end{proof}

\subsection{Theorem \ref{theo:oneshot}}
\label{proof:oneshot}

\begin{proof}
The argument is the same as in Theorem~\ref{expostprop}. Recall $\beta_{h,\mathcal Q,0}^{LP}=\lambda'(\Bar\Sigma_{\mathcal Q,0})^{-1}\Bar\Gamma_{h,0}$.
Using contemporaneous orthogonality of $\varepsilon_{x,t}$ and
$\boldsymbol\varepsilon_{z,t}$, the same block inversion yields
\[
\beta_{h,\mathcal Q,0}^{LP}
=
\theta_{y,h}-\boldsymbol\theta_z(\mathcal Q)'\Sigma_{\mathcal Q,0}^{-1}\Gamma_{h,0}.
\]
Under the linear moving-average envelope and the definition of impulse responses, $\Psi_0(\mathcal Q)'=\Sigma_{z,0}^{-1}\Sigma_{\mathcal Q,0}$ and
$\boldsymbol\psi_{y,h,0}=\Sigma_{z,0}^{-1}\Gamma_{h,0}$ with
$\Sigma_{z,0}:=\mathbb E[\boldsymbol\varepsilon_{z,t}\boldsymbol\varepsilon_{z,t}']$.
Then $\Sigma_{\mathcal Q,0}^{-1}\Gamma_{h,0}=\Psi_0(\mathcal Q)^{-1'}\boldsymbol\psi_{y,h,0}$.
In the one-shot peg, $\Delta_0^{peg}$ solves
$\Psi_0(\mathcal Q)\Delta_0^{peg}=-\boldsymbol\theta_z(\mathcal Q)$, hence
\[
\beta_{h,\mathcal Q,0}
=
\theta_{y,h}+\boldsymbol\psi_{y,h,0}'\Delta_0^{peg}
=
\theta_{y,h}-\boldsymbol\theta_z(\mathcal Q)'\Psi_0(\mathcal Q)^{-1'}\boldsymbol\psi_{y,h,0}
=
\beta_{h,\mathcal Q,0}^{LP}.
\]

For the path response, LP gives $\boldsymbol\gamma_{h,\mathcal Q,0}^{LP}=\Sigma_{\mathcal Q,0}^{-1}\Gamma_{h,0}$,
so $\boldsymbol\gamma_{h,\mathcal Q,0}^{LP}=\Psi_0(\mathcal Q)^{-1'}\boldsymbol\psi_{y,h,0}$.
The one-shot intervention solves $\Psi_0(\mathcal Q)\Delta_0^{path}=\boldsymbol c_{\mathcal Q}$, hence
$\delta_{h,0}(\boldsymbol c_{\mathcal Q})=\boldsymbol\psi_{y,h,0}'\Delta_0^{path}
=(\boldsymbol\gamma_{h,\mathcal Q,0}^{LP})'\boldsymbol c_{\mathcal Q}
=\delta_{h,0}^{LP}(\boldsymbol c_{\mathcal Q})$.
\end{proof}

\bibliographystyle{abbrvnat}
\bibliography{references}

\end{document}